\documentclass[10pt]{article}
\usepackage{algorithm}
\usepackage{bbm}
\usepackage{graphicx}
\usepackage{booktabs}
\usepackage{abbreviations}
\usepackage{lindsten}

\usepackage{color}
\renewcommand\mid{\,|\,}
\renewcommand\setX{\reals^{n_x}}
\renewcommand\setY{\reals^{n_y}}
\renewcommand\wrt{with respect to\@\xspace}
\newcommand{\rbfs}{\mbox{RB-KS}\xspace}
\newcommand{\rbffjbs}{\mbox{RB-FF/JBS}\xspace}
\newcommand{\rbffbs}{\mbox{RB-FFBS}\xspace}
\newcommand{\nls}{u}
\newcommand{\ls}{z}
\newcommand{\f}{f}
\newcommand{\g}{g}
\newcommand{\h}{h}
\newcommand{\A}{A}
\newcommand{\B}{B}
\newcommand{\C}{C}
\newcommand{\F}{F}
\newcommand{\G}{G}

\renewcommand{\Q}{Q}

\renewcommand\Mp{M}
\newcommand\Rmcmc{R}
\newcommand\bw{\widetilde w} 
\newcommand\bpw{\widetilde v} 
\newcommand\bx{\widetilde x} 
\newcommand\bnls{\widetilde \nls} 
\newcommand\bls{\widetilde \ls} 
\newcommand\bP{\widetilde P} 
\newcommand\mhalf{{-}\textstyle\frac{1}{2}} 
\newcommand\errls{ m }

\newcommand\gsmatrix{ \bar Q^{\ls} }
\newcommand\QQ{\mathcal{Q}}
\newcommand\RR{\mathcal{R}}
\newcommand\nlfunc{\varphi}

\newtheorem{lemma}{Lemma}
\newtheorem{modeldefinition}{Model}
\newtheorem{remark}{Remark}

\graphicspath{{../figs/}}

\begin{document}

\author{Fredrik Lindsten, Pete Bunch, Simo S\"arkk\"a, \\  Thomas B. Sch\"on, and Simon J. Godsill}

\date{23 May 2015}

\title{Rao-Blackwellized particle smoothers for conditionally linear Gaussian models\thanks{Supported by the projects \emph{Learning of complex dynamical systems} (Contract number:
   637-2014-466) and \emph{Probabilistic modeling of dynamical systems} (Contract number: 621-2013-
   5524), both funded by the Swedish Research Council, and the project \emph{Bayesian Tracking and Reasoning over Time} (Reference: EP/K020153/1), funded by the EPSRC.}
}
\maketitle





\begin{abstract}
Sequential Monte Carlo (\smc) methods, such as the particle filter, are by now one of the
standard computational techniques for addressing the filtering problem in general state-space models.
However, many applications 
require post-processing of data offline.
In such scenarios the smoothing problem---in which all the available data is used to compute state estimates---%
is of central interest. We consider the smoothing problem for a class of \emph{conditionally linear Gaussian}
models. We present a forward-backward-type Rao-Blackwellized particle smoother (\rbps) that is able to
exploit the tractable substructure present in these models.
Akin to the well known Rao-Blackwellized particle filter, the proposed \rbps marginalizes out a conditionally tractable
subset of state variables, effectively making use of \smc only for the ``intractable part'' of the model.
Compared to existing \rbps, two key features of the proposed method are: \textbf{\emph{(i)}} it does
not require structural approximations of the model, and \textbf{\emph{(ii)}} the aforementioned marginalization is
done both in the forward direction and in the backward direction.
\end{abstract}

\section{Introduction}\label{sec:intro}
State-space models (\ssm{s}) comprise one of the most important model
classes in statistical signal processing, automatic control, econometrics, and related areas.
A general discrete-time \ssm is given by 
\begin{subequations}
  \label{eq:intro_ssm}
  \begin{align}
    x_{t+1} &\sim p(x_{t+1} \mid x_t), \\
    y_t &\sim p(y_t \mid x_t),
  \end{align}
\end{subequations}
where $x_t\in\setX$ is the latent state process and $y_t\in\setY$ is the observed measurement process
(we use the common convention that $p$ denotes an arbitrary probability density function (\pdf)
induced by the model \eqref{eq:intro_ssm}, which is identified by its arguments).
When the model is linear and Gaussian the filtering and smoothing problems can be solved
optimally 
by using methods such as the Kalman filter 
and the Rauch-Tung-Striebel 
smoother, respectively (see, \eg, \cite{KailathSH:2000}). When going beyond the linear Gaussian case, however, no analytical solution for the optimal state inference problem is available, which calls for approximate computational methods.

Many popular deterministic methods are based on Gaussian approximations,
for instance through linearization and related techniques.
An alternative approach, for which the accuracy of the approximation is limited basically only by the computational
budget, is to use Monte Carlo methods. Among these, sequential Monte Carlo (\smc) methods such as
particle filters (PF) and particle smoothers (\ps)  play a prominent role (see, \eg, \cite{DoucetJ:2011,Gustafsson:2010a}).

While \smc can be applied directly to the general model \eqref{eq:intro_ssm}, it has been recognized that, in many cases,
there is a tractable substructure available in the model.
This structure can then be exploited to improve the performance of the \smc method.
In particular, the Rao-Blackwellized PF (\rbpf) \cite{SchonGN:2005,ChenL:2000} 
has been found to be very useful for addressing the filtering problem in \emph{conditionally linear Gaussian}
(\clg) \ssm{s} (see Section~\ref{sec:models}).
As pointed out in \cite{CappeMR:2005}, \clg models have found an ``exceptionally broad range of applications''.

However, many of the applications, as well as system identification, of \ssm{s} rely on batch analysis of data.
The central object of interest is then the smoothing distribution, that is, the
distribution of the system state(s) conditionally on all the observed data.
While there exist many \smc-based smoothers (see \eg, \cite{LindstenS:2013} and the references therein)
variance reduction by Rao-Blackwellization has not been as well explored for smoothing as for filtering.

In this paper, we present a Rao-Blackwellized \ps (\rbps) for general \clg models. The proposed method is based on the forward
filter/backward simulator (\ffbs) \cite{GodsillDW:2004}.
Contrary to the related forward-backward-type \rbps used by \cite{Kim:1994}, the
proposed method does not require any structural approximations of the model.
Another key feature of the proposed method is that it employs Rao-Blackwellization
both in the forward and backward directions, as opposed to \cite{FongGDW:2002} who
sample the full system state in the backward direction.
The use of Rao-Blackwellization also in the backward direction is necessary for the smoother to be truly Rao-Blackwellized.
An alternative \rbps, specifically targeting the marginal smoothing distribution, which is
based on the generalized two-filter formula is presented in~\cite{BriersDM:2010}.



This contribution builds upon two previous conference publications, \cite{SarkkaBG:2012} and \cite{LindstenBGS:2013},
where we studied two specific model classes, hierarchical models and mixed linear/nonlinear models, respectively (see
the next section for definitions).
Furthermore, independently of \cite{SarkkaBG:2012},
Whiteley et al. \cite{WhiteleyAD:2010} have derived essentially the same RBPS
for hierarchical models as we present here, although they study explicitly
the special case of jump Markov systems.
The present work goes beyond \cite{SarkkaBG:2012,LindstenBGS:2013} on several accounts.
First, the techniques used for the derivations and, as an effect, details of the algorithmic specifications
in these two proceedings differ substantially. Here, we harmonise the derivation and
provide a general algorithm which is applicable to both types of \clg models under study.
We also improve the previous results by extending the method to more general models. Specifically, we
allow for correlation between the process noises entering the conditionally linear and the nonlinear parts of the model,
and rank-deficient process noise covariances in the conditionally linear parts.
This comprises an important class of models in practical applications \cite{GustafssonGBFJKN:2002}.
Finally, we provide several extensions to the main method (Section~\ref{sec:extensions})
that we view as a key part of the proposed methodology.

For a vector $\mu$ and a positive semidefinite matrix $\Omega\succeq0$, we write $\| \mu \|^2_{\Omega} \eqdef \mu^\+ \Omega \mu$.
We write $|A|$ for matrix determinant and $\N(\mu, \Sigma)$ and $\N(x; \mu, \Sigma)$ for the Gaussian distribution and
\pdf, 
respectively.

\section{Conditionally linear Gaussian models}\label{sec:models}
Let the system state be partitioned into two parts: $x_t = (\nls_t, \ls_t)$, where
$\nls_t \in \reals^{n_\nls}$ is referred to as the \emph{nonlinear state} and $\ls_t \in \reals^{n_\ls}$ is referred to as the \emph{linear state}.
The \ssm \eqref{eq:intro_ssm} is said to be \clg if the conditional process $\{\ls_t, y_t \mid \nls_{1:t}\}_{t \geq 1}$ follows a time-inhomogeneous linear Gaussian
\ssm. For concreteness, we will study two specific classes of \clg models, which are of particular practical interest.
However, by combining these two model classes the proposed method can straightforwardly be generalized to other \clg models.

\begin{modeldefinition}[Hierarchical \clg model]\label{model:1}%
  A hierarchical \clg model is given by,
  \begin{subequations}
    \label{eq:hiermodel}
    \begin{align}
      \label{eq:hiermodel_a}
      \nls_{t+1} &\sim p(\nls_{t+1} \mid \nls_{t}),\\
      \label{eq:hiermodel_b}
      \ls_{t+1} &= \f(\nls_{t+1}) + \A(\nls_{t+1})\ls_{t} + \F(\nls_{t+1}) v_t, \\
      \label{eq:hiermodel_c}
      y_{t} &= \h(\nls_t) + \C(\nls_t)\ls_t + e_t,
    \end{align}
  \end{subequations}
  with process noise $v_t \sim \N(0, I_{n_v})$ and measurement noise $e_t \sim\N(0,R(\nls_t))$, respectively,
  where $R(\nls_t)$ is a positive definite matrix for any $\nls_t \in \reals^{n_\nls}$.
\end{modeldefinition}

Model~\ref{model:1} can be seen as a generalization of a
\emph{jump Markov system}, in which the ``jump'' or ``mode'' variable $\nls_t$ is allowed to be continuous.
%
However, the hierarchical structure of Model~\ref{model:1} can sometimes be limiting. We
will therefore study also the following model class.

\begin{modeldefinition}[Mixed linear/nonlinear model]\label{model:2}%
  A mixed linear/nonlinear \clg model is given by,
  \begin{subequations}
    \label{eq:mixedmodel}
    \begin{align}
      \label{eq:mixedmodel_a}
      \begin{bmatrix}
        \nls_{t+1} \\ \ls_{t+1}
      \end{bmatrix} &=
      \begin{bmatrix}
        \g(\nls_t) \\ \f(\nls_t)
      \end{bmatrix} +
      \begin{bmatrix}
        \B(\nls_t) \\ \A(\nls_t)
      \end{bmatrix}\ls_t +
      \begin{bmatrix}
        \G(\nls_t) \\ \F(\nls_t)
      \end{bmatrix}v_t, \\
      \label{eq:mixedmodel_b}
      y_{t} &= \h(\nls_t) + \C(\nls_t)\ls_t + e_t,
    \end{align}
  \end{subequations}
  with process noise $v_t \sim \N(0, I_{n_v})$ and measurement noise $e_t \sim\N(0,R(\nls_t))$, respectively,
  where $\Q(\nls_t) \eqdef \G(\nls_t)\G(\nls_t)^\+$ and $R(\nls_t)$ are assumed to be positive definite matrices for any $\nls_t \in \reals^{n_u}$.
\end{modeldefinition}

Note that Model~\ref{model:2} allows for a cross-dependence in the dynamics of the two state-components, that is
$\nls_{t+1}$ depends explicitly on $\ls_t$ and \emph{vice versa}.
Mixed linear/nonlinear models arise, for instance, when the observations
depend nonlinearly on a subset of the states in a system with linear dynamics.
See \cite{GustafssonGBFJKN:2002} for several examples from target tracking where this model is used.

\begin{remark}
We do not assume that $\F(\nls_t)\F(\nls_t)^\+$ is full rank, that is, it is only the
part of the process noise that enters on the nonlinear state $\nls_t$ that is assumed to be non-degenerate.
Note also that the model \eqref{eq:mixedmodel_a} readily allows
for correlation between the components of the process noise entering on the nonlinear state and on the linear state,
respectively.
\end{remark}

It is worth to emphasize that Model~\ref{model:1} is \emph{not} a special case of Model~\ref{model:2},
since $p(\nls_{t+1} \mid \nls_t)$ may be non-Gaussian in \eqref{eq:hiermodel_a}. Nevertheless,
Model~\ref{model:1} is simpler than Model~\ref{model:2} in many respects. Indeed, one reason for why we study
both model classes in parallel is to more clearly convey the idea of the derivation. This is possible since we can
start by looking at the (simpler) hierarchical \clg model, before generalizing the expressions to the (more involved)
mixed linear/nonlinear model.

\section{Background}

\subsection{Particle filtering and smoothing}
Consider first the general \ssm \eqref{eq:intro_ssm}.
A PF is an \smc algorithm used to approximate the intractable filtering density $p(x_t \mid y_{1:t})$ (see \eg \cite{DoucetJ:2011,Gustafsson:2010a}).
Rather than targeting the sequence of filtering densities directly, however, the PF
targets the sequence of joint smoothing densities $p(x_{1:t} \mid y_{1:t})$ for $t = 1,\,2,\,\dots$.
This is done by representing $p(x_{1:t} \mid y_{1:t})$ with a set of weighted particles $\{x_{1:t}^i, w_t^i\}_{i=1}^N$,
each of which is a state trajectory $x_{1:t}$. These particles define the point-mass approximation,
\begin{align}
  \label{eq:bkg_pfpointmass}
  \widehat p^\Np (x_{1:t} \mid y_{1:t}) \eqdef \sum_{i=1}^\Np w_{t}^i \delta_{x_{1:t}^i}(x_{1:t})     ,
\end{align}
where $\delta_x$ denotes a Dirac distribution at point $x$.
In the simplest particle filter, the $t$-th set of particles are formed by sampling $x_{1:t-1}$ from
the previous distribution (resampling)
 and then $x_{t}$ from an importance distribution $q(x_t \mid x_{1:t-1}, y_t)$.
A weight is assigned to each particle to account for the discrepancy between the proposal and the target density.
The importance weight is given by the ratio of target and proposal densities, which simplifies to,
\begin{align}
  w_t(x_{1:t}) \propto \frac{p(y_t \mid x_t) p(x_t \mid x_{t-1})}{q(x_t \mid x_{1:t-1}, y_t)}     .
\end{align}
Note that an approximation to $p(x_t \mid y_{1:t})$ is obtained by marginalization of \eqref{eq:bkg_pfpointmass},
which equates to simply discarding $x_{1:t-1}^i$ for each particle $i = \range{1}{\Np}$.

The term ``smoothing'' encompasses a number of related inference problems. Basically, it amounts to computing
the posterior \pdf of some (past) state variable, given a batch of measurements $y_{1:\T}$.
Here we focus on the estimation of the complete joint smoothing density, $p(x_{1:\T} \mid y_{1:\T})$.
Any marginal smoothing density can be computed from the joint smoothing density by marginalization.

In fact, the joint smoothing distribution is approximated at the final step of the particle filter \cite{Kitagawa:1996}.
However, this approximation suffer from the problem of path degeneracy, that is, the number of unique particles decreases rapidly for $t \ll T$ \cite{GodsillDW:2004,DoucetJ:2011}.
To mitigate this issue, a diverse set of particles may be generated by sampling state trajectories using the forward filtering/backward simulation (\ffbs) algorithm~\cite{GodsillDW:2004}.
\ffbs exploits a sequential factorization of the joint smoothing density:
\begin{align}
  p(x_{1:\T} \mid y_{1:\T}) = p(x_{\T} \mid y_{1:\T}) \prod_{t=1}^{\T-1} p(x_t \mid x_{t+1:\T}, y_{1:\T}).
\end{align}
At time $\T$, a final state $\bx_T$ is first sampled from the particle filter approximation $\widehat p^\Np (x_{T} \mid y_{1:T})$.
Then, working backward from time $\T$, each subsequent state $\bx_t$ is sampled (approximately) from the \emph{backward kernel},
$p(x_t \mid \bx_{t+1:\T}, y_{1:\T})$. The resulting trajectory $\bx_{1:\T}$ is then an approximate sample from the joint smoothing distribution.

Using the Markov property,
the backward kernel may be expressed as
\begin{align}
  \label{eq:bkg_bwdkernel_markov}
  p(x_t \mid x_{t+1:\T}, y_{1:\T}) \propto p(x_{t+1} \mid x_t) p(x_{t} \mid y_{1:t})     .
\end{align}
By using the PF approximation of the filtering distribution, we obtain the following point-mass approximation
of the backward kernel:
\begin{align}
  \label{eq:bkg_bwdkernel_pmapprox}
  \widehat p^\Np (x_{t} \mid \bx_{t+1:\T}, y_{1:\T}) \eqdef \sum_{i=1}^\Np \bw_{t|\T}^i \delta_{x_{t}^i}(x_{t}),
\end{align}
with $\bw_{t|\T}^i \propto w_t^i \thinspace p(\bx_{t+1} \mid x_{t}^i)$. The \ffbs algorithm samples
from this approximation in the backward simulation pass.

Typically, we repeat the backward simulation, say, $\Mp$ times. This generates a collection of backward trajectories
$\{\bx_{1:\T}^j\}_{j=1}^\Mp$ which define a point-mass approximation
of the joint smoothing distribution according to,
\begin{align}
  \widetilde p^\Mp (x_{1:\T} \mid y_{1:\T}) \eqdef \frac{1}{\Mp} \sum_{j=1}^\Mp \delta_{\bx_{1:\T}^j}(x_{1:\T}).
\end{align}
From this, any marginal or fixed-interval smoothing distribution can be approximated by simply discarding the
parts of the backward trajectories which are not of interest. 

\subsection{Rao-Blackwellized particle filter}\label{sec:bkg_rbpf}
We now turn our attention specifically to \clg models, such as Model~\ref{model:1} or Model~\ref{model:2}.
The structure inherent in these models can be exploited when addressing the
filtering problem. This is done in the \rbpf, which
is based on the factorization
$p(\nls_{1:t}, \ls_t \mid y_{1:t}) = p(\ls_t \mid \nls_{1:t}, y_{1:t}) p(\nls_{1:t} \mid y_{1:t})$.
Since the model is \clg, it holds that
\begin{align}
  \label{eq:bkg_rbpf_ls}
  p(\ls_t \mid \nls_{1:t}, y_{1:t}) = \N(\ls_t ; \bar\ls_{t|t}(\nls_{1:t}), P_{t|t}(\nls_{1:t}))     ,
\end{align}
for some mean and covariance functions, $\bar\ls_{t|t}(\nls_{1:t})$ and $P_{t|t}(\nls_{1:t})$, respectively.
A PF is used to estimate only the nonlinear state marginal density while conditional Kalman filters,
one for each particle, are used to compute the moments for the linear state in \eqref{eq:bkg_rbpf_ls}.
%
%
The resulting \rbpf approximation is given by
\begin{align*}
  \widehat p^\Np(\nls_{1:t}, \ls_t \mid y_{1:t}) = \sum_{i=1}^\Np w_t^i \,\N(\ls_t ; \bar\ls_{t|t}^i, P_{t|t}^i)
  \delta_{\nls_{1:t}^i}(\nls_{1:t}),
\end{align*}
where $ \bar\ls_{t|t}^i =  \bar\ls_{t|t}(\nls_{1:t}^i)$ and $ P_{t|t}^i =  P_{t|t}(\nls_{1:t}^i)$.
The particle weights are given by the ratio of $p(y_t, \nls_t \mid \nls_{1:t-1}, y_{1:t-1})$
and the importance density. See \cite{ChenL:2000} 
for details on the implementation
for the hierarchical \clg model and \cite{SchonGN:2005} for the mixed linear/nonlinear model.
The reduced dimensionality of the particle approximation results in a reduction in variance of associated estimators \cite{LindstenSO:2011,Chopin:2004}.

For numerical stability, it is recommended to implement the conditional Kalman filters
on square-root form. That is, we propagate, \eg, the Cholesky factor $\Gamma_{t|t}(\nls_{1:t})$ of the conditional covariance matrix,
rather than the covariance matrix itself, where $\Gamma_{t|t}(\nls_{1:t})$ is such that
\begin{align}
  \label{eq:bkg_rbpf_Gamma}
  P_{t|t}(\nls_{1:t}) = \Gamma_{t|t}(\nls_{1:t})\Gamma_{t|t}(\nls_{1:t})^\+.   
\end{align}
See Section~\ref{sec:extensions:squareroot}.

\section{Rao-Blackwellized particle smoothing}\label{sec:rbps}%
We now turn to the derivation of the new \rbps. The method is an \ffbs which uses the \rbpf as a forward filter.
The novelty lies in the construction of a backward simulator which samples only the nonlinear state in the backward pass.
Difficulty arises because marginally (and conditionally on the observations) the nonlinear state process is non-Markovian. Practically, this means that the backward kernel
cannot be expressed in a simple way, as in \eqref{eq:bkg_bwdkernel_markov}. We address this difficulty
by deriving a backward recursion for a set of sufficient statistics for the backward kernel.
This backward recursion is reminiscent of the backward filter in the two-filter smoothing formula for a
linear Gaussian \ssm (see \eg, \cite[Chapter~10]{KailathSH:2000}).

The basic idea is presented in Section~\ref{sec:rbps_general}, together with the statement of a general algorithm
which samples state trajectories for the nonlinear states.
We then consider the two specific model classes, the hierarchical \clg model and the mixed linear/nonlinear model,
in Section~\ref{sec:rbps_hier} and Section~\ref{sec:rbps_mixedmodel}, respectively.
In Section~\ref{sec:smoothing_linear_states} we discuss how to compute the smoothing distribution for the linear states.

\subsection{Rao-Blackwellized backward simulation}\label{sec:rbps_general}%
We wish to derive a backward simulator for the nonlinear process $\{\nls_t\}$.
That is, the target density is $p(\nls_{1:\T} \mid y_{1:\T})$.
However, when marginalizing the linear states $\{\ls_t\}$, we introduce a dependence in the measurement
likelihood on the complete history $\nls_{1:t}$. As a consequence, we must sample complete trajectories produced by the \rbpf
when simulating the nonlinear backward trajectories;
see \cite[Chapter~4]{LindstenS:2013} for a general treatment of backward simulation in the non-Markovian setting. To
solidify the idea, note that the target density can be expressed as
\begin{align}
  \label{eq:rbps_jsdfactorization}
  p(\nls_{1:\T} \mid y_{1:\T}) = p(\nls_{1:t} \mid \nls_{t+1:\T}, y_{1:\T}) p(\nls_{t+1:\T} \mid y_{1:\T}).
\end{align}
Assume that we have run a backward simulator from time $\T$ down to time $t+1$. Hence, we have generated
a partial, nonlinear backward trajectory $\bnls_{t+1:\T}$, which is an approximate sample from $p(\nls_{t+1:\T} \mid y_{1:\T})$.
To extend this trajectory to time~$t$, we draw one of the \rbpf particles $\{\nls_{1:t}^i\}_{i=1}^\Np$
(with probabilities computed below). We then set $\bnls_{t:\T} = \{\nls_t^i, \bnls_{t+1:\T}\}$ and discard $u_{1:t-1}^i$.
This procedure is then repeated for each time $t = \range{\T-1}{1}$, resulting in a complete backward trajectory $\bnls_{1:\T}$.

To compute the backward sampling probabilities, we note that the first factor in \eqref{eq:rbps_jsdfactorization} can be expressed as,
\begin{align}
  \label{eq:rbps_bwdkernel}
  p(\nls_{1:t} \mid \nls_{t+1:\T}, y_{1:\T})
  \propto p(y_{t+1:\T}, \nls_{t+1:\T} \mid \nls_{1:t}, y_{1:t}) p(\nls_{1:t} \mid y_{1:t}).
\end{align}
The second factor in this expression can be approximated by the forward \rbpf, analogously to a standard \ffbs.
Similarly to \eqref{eq:bkg_bwdkernel_pmapprox}, this results in a point-mass approximation of the backward kernel, given by
\begin{align}
  \label{eq:rbps_bwdkernel_pmapprox}
  \widehat p^\Np (\nls_{1:t} \mid \bnls_{t+1:\T}, y_{1:\T})=  \sum_{i=1}^\Np \bw_{t|\T}^i \delta_{\nls_{1:t}^i}(\nls_{1:t}),
\end{align}
with
\begin{align}
  \label{eq:rbps_weights}
  \bw_{t|\T}^i \propto w_t^i \thinspace p(y_{t+1:\T}, \bnls_{t+1:\T} \mid \nls_{1:t}^i, y_{1:t}).
\end{align}
We thus employ the following backward simulation strategy to sample $\bnls_{1:\T} \stackrel{\text{approx.}}{\sim} p(\nls_{1:\T} \mid y_{1:\T})$:
  \begin{enumerate}
  \item Run a forward RBPF for times $t = \range{1}{\T}$.
  \item Sample $\bnls_\T$ with with $\Prb(\bnls_\T = \nls_\T^i) = w_\T^i$.
  \item For $t  = \T-1$ to $1$:
    \begin{enumerate}
    \item Sample $\bnls_{t}$ with $\Prb(\bnls_{t} = \nls_t^i) = \bw_{t|\T}^i$.
    \item Set $\bnls_{t:\T} = \{ \bnls_t, \bnls_{t+1:\T}\}$.
    \end{enumerate}
  \end{enumerate}
Note that Step~3a) effectively means that we simulate $\nls_{1:t}^i$ from the point-mass approximation
of the backward kernel \eqref{eq:rbps_bwdkernel_pmapprox}, discard $\nls_{1:t-1}^i$, and set $\bnls_{t} = \nls_t^i$.
More detailed pseudo-code is given in Algorithm~\ref{alg:rbps} below.

It remains to find an expression (up to proportionality) for the predictive \pdf in \eqref{eq:rbps_weights},
in order to compute the backward sampling weights.
In fact, since the model is \clg, this \pdf can be computed straightforwardly by running a conditional Kalman filter from time $t$ up to $\T$.
However, using such an approach to calculate the weights at time $t$ would require $\Np$ separate Kalman filters to run
over $T-t$ time steps, resulting in a total computational complexity scaling quadratically with~$\T$.
To avoid this, we seek a more efficient computation of the weights \eqref{eq:rbps_weights}. This is accomplished by propagating
a set of sufficient statistics backward in time, as the trajectory $\bnls_{1:\T}$ is generated.
Specifically, these statistics are computed by running a conditional backward information filter for $z_t$,
conditionally on $\bnls_{t:\T}$, $t = \T,\,\T-1,\,\dots,\,1$.
The idea stems from \cite{GerlachCK:2000}, who use the same approach for Markov chain Monte Carlo sampling in \jmls.

To see how this can be done, note that he predictive \pdf in \eqref{eq:rbps_weights} can be expressed as
\begin{align}
  \nonumber
  p(&{}y_{t+1:\T}, \nls_{t+1:\T} \mid \nls_{1:t}, y_{1:t}) \\
  \label{eq:rbps_keyquantity}
  &= \int  p(y_{t+1:\T}, \nls_{t+1:\T} \mid \ls_t, \nls_{t}) p(\ls_t \mid \nls_{1:t}, y_{1:t}) \,d\ls_t.
\end{align}
This expression is related to the factorization used in the two-filter smoothing formula. 
The second factor of the integrand is the conditional forward filtering density. This density, computed in the forward \rbpf,
is given by \eqref{eq:bkg_rbpf_ls}. Similarly, we can view the first factor of the integrand as the density
targeted by a conditional backward filter, akin to the one used in the two-filter smoothing formula.

Indeed, we will derive a conditional backward information filter for this density, and thereby show that
\begin{align}
  \label{eq:rbps_bwdpredicted}
  p(y_{t+1:\T}, \nls_{t+1:\T} \mid \ls_t, \nls_{t}) \propto Z_t \exp\left( \mhalf \left( \ls_t^\+ \Omega_t \ls_t -2\lambda_t^\+\ls_t  \right) \right),
\end{align}
where $Z_t$, $\Omega_t \succeq 0$ and $\lambda_t$ depend on $\nls_t$, but are independent of $\ls_t$, and the proportionality is \wrt $(\nls_t,\,\ls_t)$.\footnote{By proportionality \wrt some variable $x$, we mean that the constant hidden in the proportionality sign is independent of this variable.}
Note that \eqref{eq:rbps_bwdpredicted} is not a \pdf in $\ls_t$. Still, it can be instructive to think about the above expression
as a Gaussian \pdf with information vector $\lambda_t$ and information matrix $\Omega_t$. We choose to express the backward
statistics on information form since, as we shall see later, $\Omega_t$ is not necessarily invertible. The interpretation
of \eqref{eq:rbps_bwdpredicted} as a Gaussian \pdf for $\ls_t$ implicitly corresponds to the assumption of a non-informative (flat) prior on $\ls_t$.
As pointed out above, this interpretation might be useful for understanding the role of the backward statistics,
but it does not affect our derivation in any way.

As an intermediate step of the derivation, we will also show the related identity,
\begin{align}
  \label{eq:rbps_bwdupdated}
  p(y_{t:\T}, \nls_{t+1:\T} \mid \ls_t, \nls_{t}) &\propto \exp\left( \mhalf \left( \ls_t^\+ \widehat\Omega_t \ls_t -2\widehat\lambda_t^\+\ls_t  \right) \right),
\end{align}
for some $\widehat\Omega_t \succeq 0$ and $\widehat\lambda_t$ and where the proportionality is \wrt $\ls_t$.
Computing \eqref{eq:rbps_bwdupdated} given \eqref{eq:rbps_bwdpredicted} corresponds to the measurement update of the backward
information filter (the measurement $y_t$ is taken into account). Similarly, computing \eqref{eq:rbps_bwdpredicted} given
\eqref{eq:rbps_bwdupdated}, with $t$ replaced by $t-1$, corresponds to a backward prediction.

Assume for now that \eqref{eq:rbps_bwdpredicted} holds. To compute the integral \eqref{eq:rbps_keyquantity}
we make use of the following lemma. The proof is omitted for brevity, but follows straightforwardly by plugging in the expression for $z$ and carrying out the integration \wrt $\xi$.

\begin{lemma}\label{lem:mainlemma}
  Let $\xi \sim \N(0,I)$ and let $z = c+Ax+\Gamma\xi$, for some
  constant vectors $c$ and $x$ and matrices $A$ and $\Gamma$, respectively. Let $\Omega \succeq 0$ and $\lambda$ be
  a constant matrix and vector, respectively. Then
  $\E\left[\exp\left(\mhalf\left(z^\+ \Omega z - 2\lambda^\+ z\right)\right)\right] = |M|^{-1/2}\exp\left(\mhalf \gamma\right)$ with,
  \begin{multline*}
    \gamma = \|Ax\|^2_{\Omega - \Omega\Gamma M^{-1}\Gamma^\+\Omega} - 2x^\+ A^\+\left( I-\Omega\Gamma M^{-1} \Gamma^\+  \right) m 
    \\
    + \|c\|^2_\Omega - 2\lambda^\+ c - \|\Gamma^\+ m\|^2_{M^{-1}},
  \end{multline*}
  where $m = \lambda - \Omega c$ and $M = \Gamma^\+ \Omega \Gamma + I$.
\end{lemma}

From \eqref{eq:bkg_rbpf_ls} and \eqref{eq:bkg_rbpf_Gamma} it follows that if we write
\begin{align}
  \label{eq:rbps_ls_filter}
  \ls_{t} &= \bar \ls_{t|t} + \Gamma_{t|t} \xi_{t}, & \xi_{t}&\sim \N(0,I),
\end{align}
then the distribution of $\ls_t$ in \eqref{eq:rbps_ls_filter} is $p(\ls_t \mid \nls_{1:t}, y_{1:t})$.
In the above, we have dropped the dependence on $\nls_{1:t}$ for brevity.
The integral in \eqref{eq:rbps_keyquantity} can thus be computed by applying Lemma~\ref{lem:mainlemma} with
$c = \bar\ls_{t|t}$, $x = 0$, $\Gamma = \Gamma_{t|t}$, $\Omega = \Omega_{t}$ and $\lambda = \lambda_t$.
It follows that,
\begin{align}
  p(y_{t+1:\T}, \nls_{t+1:\T} \mid \nls_{1:t}, y_{1:t}) \propto Z_t |\Lambda_t|^{-1/2} \exp\left( \mhalf \eta_{t} \right),
\end{align}
where the proportionality is \wrt $\nls_{1:t}$ and with,
\begin{subequations}
  \label{eq:rbps_etalambda}
  \begin{align}
    \eta_t &= \| \bar\ls_{t|t}\|^2_{\Omega_t} - 2\lambda_t^\+ \bar\ls_{t|t} -
    \| \Gamma_{t|t}^\+ (\lambda_t - \Omega_t \bar\ls_{t|t}) \|^2_{\Lambda_t^{-1}}, \\
    \Lambda_t &= \Gamma_{t|t}^\+ \Omega_t \Gamma_{t|t} + I.
  \end{align}
\end{subequations}
By plugging this result into \eqref{eq:rbps_weights}, we obtain an expression for the backward sampling weights.
It remains to show the identity \eqref{eq:rbps_bwdpredicted} and to find the updating equations
for the statistics $\{Z_t, \Omega_t, \lambda_t\}$. These recursions will be derived explicitly for
the two model classes under study in the consecutive two sections.
The resulting Rao-Blackwellized backward simulator
is given in Algorithm~\ref{alg:rbps}. As for a standard \ffbs, the backward simulation
is typically repeated $\Mp$ times, to generate a collection of backward trajectories $\{ \bnls_{1:\T}^j \}_{j=1}^\Mp$
which can be used to approximate $p(\nls_{1:\T} \mid y_{1:\T})$.


\begin{algorithm}
  \caption{Rao-Blackwellized backward simulator}
  \label{alg:rbps}
  \begin{enumerate}
  \item \textbf{Forward filter:} Run an \rbpf for time $t = \range{1}{\T}$. For each $t$, store $\{\nls_t^i, w_t^i, \bar\ls_{t|t}^i, \Gamma_{t|t}^i\}_{i=1}^\Np$.
  \item \textbf{Initialize:} Draw $\bnls_\T$ with $\Prb(\bnls_\T = \nls_\T^i) = w_\T^i$. Compute $\widehat\Omega_\T$ and $\widehat\lambda_\T$
    according to \eqref{eq:rbps_hier_statistics_init}.
  \item \textbf{For $t = T-1$ to $1$:}
    \begin{enumerate}
    \item \textbf{Backward filter prediction:}
    \item[] \textit{(Model 1: hierarchical)}
      \begin{itemize}
      \item[-] Compute $Z_t^i = p(\bnls_{t+1} \mid \nls_t^i)$ for $i = \range{1}{N}$.
      \item[-] Compute $\{\Omega_t, \lambda_t \}$ according to \eqref{eq:rbps_hier_statistics_pred}.
      \end{itemize}
    \item[] \textit{(Model 2: mixed)}
      \begin{itemize}
      \item[-] Compute $\{Z_t^i, \Omega_t^i, \lambda_t^i \}$ according to \eqref{eq:rbps_mixed_statistics_pred}
        for each forward filter particle $i = \range{1}{N}$.
      \end{itemize}
    \item For $i = \range{1}{\Np}$:
      \begin{enumerate}
      \item Compute $\{\Lambda_t^i, \eta_t^i\}$ according to \eqref{eq:rbps_etalambda}.
      \item Compute $W_t^i = w_t^i Z_t^i |\Lambda_t^i|^{-1/2} \exp\left(\mhalf\eta_t^i\right)$.
      \end{enumerate}
    \item Normalize the weights, $\bw_{t|\T}^i = W_t^i / \sum_l W_t^l$.
    \item Draw $J$ with $\Prb(J = i) = \bw_{t|\T}^i$.
    \item Set $\bnls_{t:\T} = \{\nls_t^J, \bnls_{t+1:\T}\}$.
    \item \textit{(Model 2: mixed)} Set $\{\Omega_t, \lambda_t\} = \{\Omega_t^J, \lambda_t^J\}$.
    \item \textbf{Backward filter measurement update:} Compute $\{\widehat\Omega_{t}, \widehat\lambda_t\}$ according to \eqref{eq:rbps_hier_statistics_update}.
    \end{enumerate}
  \end{enumerate}
\end{algorithm}

\subsection{Model 1 -- Hierarchical \clg model}\label{sec:rbps_hier}%
We now consider Model~\ref{model:1}, the hierarchical \clg model,
and prove the identities \eqref{eq:rbps_bwdpredicted} and \eqref{eq:rbps_bwdupdated}.
We also derive explicit updating equations for the statistics $\{Z_t, \Omega_t, \lambda_t\}$ and $\{\widehat \Omega_t, \widehat\lambda_t\}$, respectively.

\begin{remark}
  The expressions derived in this section have previously been presented by \cite{WhiteleyAD:2010}
  who, independently from our preliminary work in \cite{SarkkaBG:2012}, have derived an \rbps for hierarchical
  \clg models. Nevertheless, we believe that the present section will
  be useful in order to make the derivation for the (more involved)
  mixed linear/nonlinear model in Section~\ref{sec:rbps_mixedmodel} more accessible.
\end{remark}


For notational simplicity, we write $\A_t$ for $\A(\nls_t)$ and similarly for other functions of $\nls_t$.
To initialize the backward statistics at time $\T$, we note that \eqref{eq:hiermodel_c} can be written as,
\begin{align}
  p(y_{\T} \mid \ls_\T, \nls_\T) = \N(y_\T ; \h_\T + \C_\T \ls_\T, R_\T)
  \propto \exp\left( \mhalf \left( \ls_\T^\+ \widehat\Omega_\T \ls_\T -2\widehat\lambda_\T^\+\ls_\T  \right) \right),
\end{align}
with
\begin{subequations}
  \label{eq:rbps_hier_statistics_init}
  \begin{align}
    \widehat\Omega_{\T} &= \C_\T^\+ R_\T^{-1} \C_\T, \\
    \widehat\lambda_\T &= \C_\T^\+ R_\T^{-1} (y_\T - \h_\T),
  \end{align}
\end{subequations}
which shows that \eqref{eq:rbps_bwdupdated} holds at time $t=\T$ (with the convention $u_{\T+1:\T} = \emptyset$).
We continue by using an inductive argument.
Hence, assume that \eqref{eq:rbps_bwdupdated} holds at some time $t+1$.
To prove \eqref{eq:rbps_bwdpredicted} we do a backward prediction step. We have, for $t < \T$,
\begin{multline}
  \label{eq:rbps_hier_bwdpred}
  p(y_{t+1:\T}, \nls_{t+1:\T} \mid \ls_t, \nls_{t})  \\ 
  = p(\nls_{t+1} \mid \nls_t)
  \int p(y_{t+1:\T}, \nls_{t+2:\T} \mid \ls_{t+1}, \nls_{t+1}) p(\ls_{t+1} \mid \ls_t, \nls_{t+1})\,d\ls_{t+1}.
\end{multline}
Using the induction hypothesis and \eqref{eq:hiermodel_b}, the above integral can be computed by applying Lemma~\ref{lem:mainlemma} with
$c = \f_{t+1}$, $A = \A_{t+1}$, $x = \ls_t$, $\Gamma = \F_{t+1}$, $\Omega = \widehat\Omega_{t+1}$
and $\lambda  = \widehat\lambda_{t+1}$. It follows that \eqref{eq:rbps_hier_bwdpred} coincides with \eqref{eq:rbps_bwdpredicted},
with
\begin{subequations}
  \label{eq:rbps_hier_statistics_pred}
  \begin{align}
    \label{eq:rbps_hier_statistics_pred_a}
    Z_t &= p(\nls_{t+1} \mid \nls_t), \\
    \label{eq:rbps_hier_statistics_pred_b}
    \Omega_{t} &= \A_{t+1}^\+ \left(I - \widehat\Omega_{t+1} \F_{t+1} M_{t+1}^{-1} \F_{t+1}^\+ \right)\widehat\Omega_{t+1} \A_{t+1}, \\
    \label{eq:rbps_hier_statistics_pred_c}
    \lambda_t &= \A_{t+1}^\+ \left(I - \widehat\Omega_{t+1} \F_{t+1} M_{t+1}^{-1} \F_{t+1}^\+ \right) \errls_{t+1},
  \end{align}
  where we have defined the quantities $\errls_{t+1} \eqdef \widehat\lambda_{t+1} - \widehat\Omega_{t+1} \f_{t+1}$
  and $M_{t+1} \eqdef \F_{t+1}^\+ \widehat\Omega_{t+1} \F_{t+1} + I$.
\end{subequations}%
Note that the above statistics depend on $\nls_t$ only through the factor $p(\nls_{t+1} \mid u_t)$ in \eqref{eq:rbps_hier_statistics_pred_a}.
This is important from an implementation point of view, since it implies that we do not need to make the backward prediction
for each forward filter particle; see Algorithm~\ref{alg:rbps}.

Next, to prove \eqref{eq:rbps_bwdupdated} for $t < \T$, we assume that \eqref{eq:rbps_bwdpredicted} holds at time $t$.
We have,
\begin{align}
  \label{eq:rbps_hier_bwdupdate}
   p(y_{t:\T}, \nls_{t+1:\T} \mid \ls_t, \nls_{t})
   =  p(y_t \mid \ls_t, \nls_t) p(y_{t+1:\T}, \nls_{t+1:\T} \mid \ls_t, \nls_{t}).
\end{align}
The first factor is given by \eqref{eq:hiermodel_c}, analogously to \eqref{eq:rbps_hier_statistics_init}, and the second factor is given by
\eqref{eq:rbps_bwdpredicted}. By collecting terms from the two factors, we see that \eqref{eq:rbps_hier_bwdupdate} coincides
with \eqref{eq:rbps_bwdupdated}, where
\begin{subequations}
  \label{eq:rbps_hier_statistics_update}
  \begin{align}
    \widehat\Omega_{t} &= \Omega_t + \C_t^\+ R_t^{-1} \C_t, \\
    \widehat\lambda_t &= \lambda_t + \C_t^\+ R_t^{-1} (y_t - \h_t).
  \end{align}
\end{subequations}
As pointed out above, this correspond to the backward measurement update. Since we are working with the information
form of the backward filter, the measurement update simply corresponds to the addition of a term to the information
vector and the information matrix, respectively.

\subsection{Model 2 -- Mixed linear/nonlinear \clg model}\label{sec:rbps_mixedmodel}%
We now turn to the mixed linear/nonlinear model \eqref{eq:mixedmodel} and prove the identities \eqref{eq:rbps_bwdpredicted} and \eqref{eq:rbps_bwdupdated}
for this class of systems. First, note that the measurement equations
are identical for the models \eqref{eq:hiermodel} and \eqref{eq:mixedmodel}. Consequently,
the initialization \eqref{eq:rbps_hier_statistics_init} and the backward measurement update \eqref{eq:rbps_hier_statistics_update}
hold for the mixed linear/nonlinear model as well. We will thus focus on the backward prediction step.

Similarly to \eqref{eq:rbps_hier_bwdpred} we factorize the backward prediction density according to,
\begin{multline}
  \label{eq:rbps_mixed_bwdpred}
  p(y_{t+1:\T}, \nls_{t+1:\T} \mid \ls_t, \nls_{t}) \\ = p(\nls_{t+1} \mid \nls_t, \ls_t)
  \int p(y_{t+1:\T}, \nls_{t+2:\T} \mid \ls_{t+1}, \nls_{t+1}) p(\ls_{t+1} \mid \ls_t, \nls_t, \nls_{t+1})\,d\ls_{t+1}.
\end{multline}
Note that the first factor now depends on $\ls_t$.
From \eqref{eq:mixedmodel_a}, we can express this density as,
\begin{multline}
  \label{eq:rbps_mixed_pu}
  p(\nls_{t+1}\mid \ls_t, \nls_t) = \N(\nls_{t+1} ; \g_t + \B_t\ls_t, Q_t) \\
  \propto |Q_t|^{-1/2} \exp\left( \mhalf \left( \|\nls_{t+1} - \g_t\|^2_{Q_t^{-1}}\right) \right) \\
  \times
  \exp\left( \mhalf \left( \| \B_t \ls_t\|^2_{Q_t^{-1}} - 2\ls_t^\+ \B_t^\+ Q_t^{-1} (\nls_{t+1} - \g_t) \right) \right).
\end{multline}
Next, we address the integral in \eqref{eq:rbps_mixed_bwdpred}.
Since the process noise $v_t$ enters the expressions for both $\nls_{t+1}$ and $\ls_{t+1}$
in \eqref{eq:mixedmodel_a}, there is a statistical dependence between
$\ls_{t+1}$ and $\nls_{t+1}$. In other words, since we allow
for cross-correlation between the process noises entering on $\nls_{t+1}$ and $\ls_{t+1}$, respectively,
knowledge about $\nls_{t+1}$ will contain information about $\ls_{t+1}$. This has to be taken into
account when computing the second factor of the integrand in \eqref{eq:rbps_mixed_bwdpred}. To handle this,
we make use of a Gram-Schmidt orthogonalization to decorrelate the process noises. Let $\bar v_t^z = \gsmatrix_t v_t$, where
\begin{align}
  \gsmatrix_t \eqdef I - G_t^\+ Q_t^{-1} G_t.
\end{align}
Note that $\gsmatrix_t$ is a projection matrix: $(\gsmatrix_t)^2 = \gsmatrix_t$.
It follows that $\E[  \bar v_t^z  (Gv_t)^\+ ] = 0$ and $\E[  \bar v_t^z  (\bar v_t^z)^\+ ] = \gsmatrix_t$,
We can then rewrite the dynamical equation \eqref{eq:mixedmodel_a} as,
\begin{subequations}
  \label{eq:rbps_mixed_orthogonalmodel}
  \begin{align}
    u_{t+1} &= \g_t + \B_t \ls_t + \G_t v_t, \\
    \label{eq:rbps_mixed_orthogonalmodel_b}
    z_{t+1} &= \bar\f_t + \bar\A_t \ls_t + \F_t \bar v_t^z,
  \end{align}
  where
  \begin{align}
    \bar f_t &= \bar\f_t(\nls_t, \nls_{t+1}) \eqdef \f_t + \F_t\G_t^\+ Q_t^{-1}(u_{t+1} -\g_t), \\
    \bar A_t &= \bar\A_t(\nls_t) \eqdef \A_t - \F_t\G_t^\+ Q_t^{-1} \B_t,
  \end{align}
\end{subequations}%
and where the process noises entering on $\nls_{t+1}$ and $\ls_{t+1}$ are now independent.
Hence, from \eqref{eq:rbps_mixed_orthogonalmodel_b}, we can write
\begin{align}
  p(\ls_{t+1} \mid \ls_t, \nls_t, \nls_{t+1}) = \N(\ls_{t+1} ; \bar\f_t + \bar\A_t \ls_t, \F_t \gsmatrix_t \F_t^\+).
\end{align}
The integral in \eqref{eq:rbps_mixed_bwdpred} can now be computed by applying Lemma~\ref{lem:mainlemma}
with $c = \bar\f_t$, $A = \bar\A_t$, $x = \ls_t$, $\Gamma = \F_t \gsmatrix_t$, $\Omega = \widehat\Omega_{t+1}$
and $\lambda  = \widehat\lambda_{t+1}$. Combining this
result with \eqref{eq:rbps_mixed_pu} and collecting the terms, we see that \eqref{eq:rbps_mixed_bwdpred} coincides with \eqref{eq:rbps_bwdpredicted} with,
\begin{subequations}
  \label{eq:rbps_mixed_statistics_pred}
  \begin{align}
    Z_t &=  |\Q_t|^{-1/2} |M_t|^{-1/2} \exp\left(\mhalf \tau_t \right),\\
    \Omega_{t} &= \bar\A_{t}^\+ \left(I - \widehat\Omega_{t+1} \F_t \Psi_t \F_{t}^\+ \right)\widehat\Omega_{t+1} \bar\A_{t} + \B_t^\+ Q_t^{-1} \B_t, \\
    \lambda_t &= \bar\A_{t}^\+ \left(I - \widehat\Omega_{t+1} \F_{t} \Psi_t \F_{t}^\+ \right) \errls_{t},
  \end{align}
  where we have defined the quantities
  \begin{align*}
    \tau_t &= \| \nls_{t+1}-\g_t \|^2_{Q_t^{-1}} + \| \bar\f_t \|^2_{\widehat\Omega_{t+1}}
     -2\widehat\lambda_{t+1}^\+ \bar\f_t -  \| \F_t^\+ \errls\|^2_{\Psi_t},\\
    \Psi_t &=   \gsmatrix_{t} M_t^{-1} \gsmatrix_{t}, \\
    M_{t} &=  \gsmatrix_{t}\F_{t}^\+ \widehat\Omega_{t+1} \F_{t} \gsmatrix_{t} + I, \\
    \errls_{t} &= \widehat\lambda_{t+1} - \widehat\Omega_{t+1} \bar\f_{t}.
  \end{align*}
\end{subequations}
As opposed to the hierarchical model, the predicted backward statistics $\{Z_t, \Omega_t, \lambda_t\}$ all depend explicitly on $\nls_t$
for this model. This implies that the backward prediction has to be done for each forward filter particle, see Algorithm~\ref{alg:rbps}.
It should be noted, however, that the updating equations \eqref{eq:rbps_mixed_statistics_pred} can be simplified
for some special cases of the mixed linear/nonlinear model \eqref{eq:mixedmodel}. In particular,
if the dynamics \eqref{eq:mixedmodel_a} are Gaussian and linear in both $\ls_t$ and $\nls_t$
(the measurement equation \eqref{eq:mixedmodel_b} may be nonlinear in $\nls_t$), it is enough to do one backward prediction.
Models with linear dynamics and nonlinear measurement equations are indeed common in many applications, see~\cite{GustafssonGBFJKN:2002}.

\subsection{Smoothing the linear states}\label{sec:smoothing_linear_states}
Algorithm~\ref{alg:rbps} provides a way of simulating nonlinear state trajectories, approximately distributed
according to \mbox{$p(\nls_{1:\T} \mid y_{1:\T})$}. However, it is often the case that we are also interested in smoothed estimates
of the linear states $\{\ls_t\}$. These estimates can be obtained by fusing the statistics from a forward conditional Kalman filter, with
the backward statistics computed during the backward simulation. Note, however, that the forward statistics need to be
computed anew; that is, we can \emph{not} simply use the statistics from
the forward \rbpf. The reason is that the statistics should be computed conditionally on the nonlinear trajectories
simulated in the backward sweep, which are in general different from the trajectories simulated by the \rbpf.

Let $\bnls_{1:\T}$ be a backward trajectory generated by Algorihtm~\ref{alg:rbps}.
To compute the conditional smoothing \pdf for $\ls_t$ we start by noting that
\begin{align}
  p(\ls_t \mid \bnls_{1:\T}, y_{1:\T}) \propto p(y_{t+1:\T}, \bnls_{t+1:\T} \mid \ls_t, \bnls_t) p(\ls_t \mid \bnls_{1:t}, y_{1:t}).
\end{align}
Since the model is \clg, the latter factor can be computed by running a Kalman filter, conditionally on the fixed nonlinear state trajectory
$\bnls_{1:t}$. We get,
\begin{align}
  p(\ls_t \mid \bnls_{1:t}, y_{1:t}) = \N(\ls_t ; \bls_{t|t}, \bP_{t|t}),
\end{align}
for some mean vector $ \bls_{t|t}$ and covariance matrix $\bP_{t|t}$, respectively (\cf, \eqref{eq:bkg_rbpf_ls}). By fusing this information
with the backward information filter, given by \eqref{eq:rbps_bwdpredicted}, we get,%
\begin{subequations}
\label{eq:rbps_linear_state_smoothing}
\begin{align}
  p(\ls_t \mid \bnls_{1:\T}, y_{1:\T}) = \N(\ls_t ; \bls_{t|\T}, \bP_{t|\T}),
\end{align}
with
\begin{align}
  \bls_{t|\T} &=  \bP_{t|\T} \left( \bP_{t|t}^{-1}\bls_{t|t} + \lambda_t \right), \\
  \bP_{t|\T} &= \left( \bP_{t|t}^{-1} + \Omega_t  \right)^{-1}.
\end{align}
\end{subequations}
The resulting method can be seen as a forward-backward-forward smoother. First, a forward \rbpf is used to filter the data. Second,
a backward simulator is applied to simulate nonlinear state trajectories. Finally, a new forward sweep is carried out
to compute the smoothing distributions for the linear states.
The complete \rbps is given in Algorithm~\ref{alg:rbps-2}.

\begin{algorithm}
  \caption{Rao-Blackwellized particle smoother (\rbps)}
  \label{alg:rbps-2}
  \begin{enumerate}
  \item \textbf{Forward filter/backward simulator:} Run Algorithm~\ref{alg:rbps} to simulate a nonlinear state trajectory $\bnls_{1:\T}$. For each $t = \range{1}{\T}$, store $\Omega_t$ and $\lambda_t$.
  \item \textbf{Linear state smoothing:}
    \begin{enumerate}
    \item Run a Kalman filter for the linear states, conditionally on $\bnls_{1:\T}$.
      For each $t = \range{1}{\T}$, store the filtered mean and covariance: $\{\bls_{t|t}, \bP_{t|t}\}$.
      \item Compute the smoothed means and covariances according to \eqref{eq:rbps_linear_state_smoothing}.
    \end{enumerate}
  \end{enumerate}
\end{algorithm}

Similarly to above, we may also compute, for instance, the two-step smoothing distribution
$ p(\ls_{t-1:t} \mid \bnls_{1:\T}, y_{1:\T})$ which is typically required when using the
smoother for parameter estimation.

\section{Extensions and computational aspects}\label{sec:extensions}%

\subsection{Approximate Rao-Blackwellization}\label{sec:extensions:approx}%
As pointed out in Section~\ref{sec:intro}, an alternative to \smc is to use some
deterministic Gaussian approximation of the filtering and smoothing
distributions. This gives rise to methods such as the extended and the unscented
Kalman filters and smoothers.
In \cite{WanM:2001}, an unscented two-filter smoother is constructed by inverting the dynamical model.
However, as pointed out in \cite{Briers:2007}, inversion of the dynamics will in general not lead to the correct result.
Instead, \cite{Briers:2007} suggest a generalized two-filter smoothing formula and use this as a basis for an
unscented two-filter smoother (see also \cite{BriersDM:2010}).

It is possible to combine these methods with the proposed \rbps. This enables
smoothing for general nonlinear state-space models, in which one part of the state vector
is approximated using particles and the other part of the state vector
is handled using a deterministic approximation. This hybrid approach can be useful
when deterministic approximations are found to be appropriate for some state variables,
but insufficient for some other variables. 

Consider the following, general nonlinear \ssm,%
\begin{subequations}
  \label{eq:extensions_approx_model}
  \begin{align}
    \label{eq:extensions_approx_model_a}
    x_{t+1} = 
    \begin{pmatrix}
      \nls_{t+1} \\ \ls_{t+1}
    \end{pmatrix} &= \f(\nls_t, \ls_t, v_t), \\
    \label{eq:extensions_approx_model_b}
    y_{t} &= \h(\nls_t, \ls_t, e_t).
  \end{align}
\end{subequations}
with process noise $v_t \sim \N(0, I_{n_v})$ and measurement noise $e_t \sim\N(0,I_{n_e})$, respectively.
The partitioning of the state according to $x_t = (\nls_t, \ls_t)$ is in this case superficial, since
the model is nonlinear in both variables. However, the partitioning is used to indicate which part of the model that we intend
to address using particles, and which part that we intend to address using a deterministic approximation.

Approximate Rao-Blackwellized forward filtering can be done for the model \eqref{eq:extensions_approx_model}
by using, for instance, an extended or an unscented \rbpf \cite{SarkkaVL:2007}. These methods
are based on different types of Gaussian approximations.
Let $x \in \reals^{n_x}$ be a Gaussian distributed random vector and let $\nlfunc: \reals^{n_x} \rightarrow \reals^{n_z}$ be
some (nonlinear) transformation. A Gaussian approximation scheme can be used to find a Gaussian approximation of the random vector $z = \nlfunc(x)$.
Examples of such approximations are first and second order Taylor expansions, \ie linearizations,
and the unscented transform \cite{JulierU:2004}.
Assume that the forward filter \eqref{eq:bkg_rbpf_ls} holds
approximately. We then seek a generalization of the backward information filter given by \eqref{eq:rbps_bwdpredicted}
and \eqref{eq:rbps_bwdupdated} to the nonlinear setting.
We suggest an approach which draws upon the generalized two-filter smoothing formula by \cite{BriersDM:2010,Briers:2007}.

Consider first the backward prediction step \eqref{eq:rbps_mixed_bwdpred}.
Let us introduce the auxiliary quantities
\begin{align}
  \label{eq:extensions_approx_gammadef}
  \gamma_t(\ls_t) \eqdef \N(\ls_t ; \mu_t, \Sigma^\ls_t),
\end{align}
for some user-chosen parameters $\mu_t$ and $\Sigma^\ls_t$. In \cite{BriersDM:2010,Briers:2007}, these
functions are viewed as \emph{artificial priors}. Indeed, if \eqref{eq:extensions_approx_gammadef} is
viewed as a prior distribution on $\ls_t$, then \eqref{eq:extensions_approx_model_a} is a nonlinear
transformation of the Gaussian vector ($\nls_t$ is fixed),
\begin{align}
  \label{eq:extensions_approx_zv}
  \begin{pmatrix}
    \ls_t \\ v_t
  \end{pmatrix} \sim \N\left(
    \begin{pmatrix}
      \mu_t \\ 0
    \end{pmatrix},
    \begin{pmatrix}
      \Sigma^\ls_t & 0 \\
      0 & I_{n_v}
    \end{pmatrix}\right).
\end{align}
To exploit this, we write \eqref{eq:rbps_mixed_bwdpred} as
\begin{multline}
  \label{eq:extensions_approx_dynamics}
  p(y_{t+1:\T}, \nls_{t+1:\T} \mid \ls_t, \nls_{t}) \\
  = \int p(y_{t+1:\T}, \nls_{t+2:\T} \mid \ls_{t+1}, \nls_{t+1}) \frac{\widetilde p(x_{t+1}, \ls_t \mid \nls_t)}{\gamma_t(\ls_t)} \,d\ls_{t+1}.
\end{multline}
with $\widetilde p(x_{t+1}, \ls_t \mid \nls_t) \eqdef  p(\ls_{t+1}, \nls_{t+1} \mid \ls_t, \nls_t) \gamma_{t}(\ls_t)$.
By using \eqref{eq:extensions_approx_zv} and applying a Gaussian approximation scheme to the mapping,
\begin{align}
  \begin{pmatrix}
    x_{t+1} \\ \ls_t
  \end{pmatrix} =
  \begin{pmatrix}
    f(\nls_t, \ls_t, v_t) \\ \ls_t
  \end{pmatrix},
\end{align}
we get
\begin{align}
  \label{eq:extensions_approx_garesult}
  \begin{pmatrix}
    x_{t+1} \\ \ls_t
  \end{pmatrix} \stackrel{\text{approx.}}{\sim} \N\left(
    \begin{pmatrix}
      c_t \\ \mu_t
    \end{pmatrix},
    \begin{pmatrix}
      \Sigma^x_t & \Sigma^{x\ls}_t \\ (\Sigma^{x\ls}_t)^\+ & \Sigma^\ls_t
    \end{pmatrix}\right),
\end{align}
for some vector $c_t$ and matrices $\Sigma^{x}_t$ and $\Sigma^{x\ls}_t$, respectively.
Here, we have made the nonrestrictive assumption that the Gaussian approximation scheme
applied to the identity mapping retains the Gaussian prior.
By factorizing \eqref{eq:extensions_approx_garesult} we have%
\begin{subequations}
  \label{eq:extensions_approx_dynamicsapproximation}
  \begin{align}
    \widetilde p(x_{t+1}, \ls_t \mid \nls_t) \approx \N(x_{t+1} ; \f^{x}_t + \A^{x}_t \ls_t, Q^{x}_t ) \gamma_t(\ls_t),
  \end{align}
  where
  \begin{align}
    \label{eq:extensions-Ax}
    \A^{x}_t &= \Sigma^{x\ls}_t (\Sigma^\ls_t)^{-1}, \\
    \label{eq:extensions-fx}
    \f^{x}_t &= c_t - \A^{x}_t \mu_t, \\
    \label{eq:extensions-Qx}
     Q^{x}_t &= \Sigma^x_t - \A^{x}_t (\Sigma^{x\ls}_t)^\+.
  \end{align}
\end{subequations}
By plugging \eqref{eq:extensions_approx_dynamicsapproximation} into \eqref{eq:extensions_approx_dynamics}, we see that
the factor $\gamma_t(\ls_t)$ cancels. We can thus use the approximate dynamics defined by \eqref{eq:extensions_approx_dynamicsapproximation}
in the updating formulas for the backward prediction \eqref{eq:rbps_mixed_statistics_pred}. To recover the notation used in
\eqref{eq:rbps_mixed_orthogonalmodel} and \eqref{eq:rbps_mixed_statistics_pred},
however, we need to split the quantities defined in \eqref{eq:extensions_approx_dynamicsapproximation} according
to the two state components $\nls_{t+1}$ and $\ls_{t+1}$, respectively.
That is, we define $\g_t$, $\f_t$, $\B_t$, $\A_t$, $\G_t$ and $\F_t$ through,
\begin{align*}
  \f^x_t &=
  \begin{pmatrix}
    \g_t \\ \f_t
  \end{pmatrix}, &
  \A^x_t &=
  \begin{pmatrix}
    \B_t \\ \A_t
  \end{pmatrix}, &
  Q^x_t &=
  \begin{pmatrix}
    \G_t \\ \F_t
  \end{pmatrix}
  \begin{pmatrix}
    \G_t \\ \F_t
  \end{pmatrix}^\+,
\end{align*}
where the latter expression is given by for instance a Cholesky factorization of $Q^x_t$.

The backward measurement update \eqref{eq:rbps_hier_bwdupdate} can be handled in a similar way. We write \eqref{eq:rbps_hier_bwdupdate}
as
\begin{align}
  \label{eq:extensions_approx_update}
   p(y_{t:\T}, \nls_{t+1:\T} \mid\ls_t, \nls_{t}) 
   = p(y_{t+1:\T}, \nls_{t+1:\T} \mid \ls_t, \nls_{t}) \frac{\widetilde p(y_t, \ls_t \mid \nls_t)}{\gamma_t(\ls_t)},
\end{align}
with $\widetilde p(y_t, \ls_t \mid \nls_t) \eqdef p(y_t \mid \ls_t, \nls_t) \gamma_t(\ls_t)$ and $\gamma_t(\ls_t)$
being a user-chosen Gaussian density as in \eqref{eq:extensions_approx_gammadef}, possibly different from the one
used in the prediction step. As above, with $\gamma_t(\ls_t)$ interpreted as an artificial prior, \eqref{eq:extensions_approx_model_b}
is a nonlinear transformation of the Gaussian vector,
\begin{align}
  \label{eq:extensions_approx_ze}
  \begin{pmatrix}
    z_t \\ e_t
  \end{pmatrix} \sim \N\left(
    \begin{pmatrix}
      \mu_t \\ 0
    \end{pmatrix},
    \begin{pmatrix}
      \Sigma^\ls_t & 0 \\
      0 & I_{n_e}
    \end{pmatrix}\right).
\end{align}
By applying a Gaussian approximation scheme to the mapping,
\begin{align}
  \begin{pmatrix}
    y_t \\ z_t
  \end{pmatrix} =
  \begin{pmatrix}
    h(\nls_t, \ls_t, e_t) \\ z_t
  \end{pmatrix},
\end{align}
we get
\begin{align}
  \label{eq:extensions_approx_garesult2}
  \begin{pmatrix}
    y_t \\ \ls_t
  \end{pmatrix} \stackrel{\text{approx.}}{\sim} \N\left(
    \begin{pmatrix}
      d_t \\ \mu_t
    \end{pmatrix},
    \begin{pmatrix}
      \Sigma^y_t & \Sigma^{y\ls}_t \\ (\Sigma^{y\ls}_t)^\+ & \Sigma^\ls_t
    \end{pmatrix}\right),
\end{align}
for some vector $d_t$ and matrices $\Sigma^{y}_t$ and $\Sigma^{y\ls}_t$, respectively.
By factorizing \eqref{eq:extensions_approx_garesult2} we have%
\begin{subequations}
  \begin{align}
    \widetilde p(y_t, \ls_t \mid \nls_t) \approx \N(y_t ; \h_t + \C_t \ls_t, R_t ) \gamma_t(\ls_t),
  \end{align}
  where
  \begin{align}
    \C_t &= \Sigma^{y\ls}_t (\Sigma^\ls_t)^{-1}, \\
    \h_t &= d_t - \C_t \mu_t, \\
     R_t &= \Sigma^y_t - \C_t (\Sigma^{y\ls}_t)^\+.
  \end{align}
\end{subequations}
The above quantities can then be used in the backward measurement update equations \eqref{eq:rbps_hier_statistics_update}.

To apply the approximate \rbps as described above, we need to choose the artificial priors \eqref{eq:extensions_approx_gammadef}.
In \cite{BriersDM:2010}, it is suggested to use the actual ``prior'' $\gamma_t(\ls_t) = p(\ls_t \mid \nls_t)$
 (recall that $\nls_t$ is fixed at this stage of the algorithm),
or some approximation of this density. However, it is also pointed out that the approach is more generally applicable.
Indeed, requiring $\gamma_t(\ls_t)$ to be close to $p(\ls_t \mid \nls_t)$ is only important if we want
$\widetilde p(x_{t+1}, \ls_t \mid \nls_t)$ and $\widetilde p(y_t, \ls_t \mid \nls_t)$ to be close approximations
to $p(x_{t+1}, \ls_t \mid \nls_t)$ and $p(y_t, \ls_t \mid \nls_t)$, respectively. For our purposes, this is not
necessary, since the artificial priors cancel in \eqref{eq:extensions_approx_dynamics} and \eqref{eq:extensions_approx_update}.
In fact, $\gamma_t(\ls_t)$ serves as a type of indicator for the operational range in the state-space of the Gaussian approximations.
A more natural choice might thus be to use the current estimate of $\ls_t$ to specify $\gamma_t(\ls_t)$,
extracted either from the forward filter of from the backward filter.
Indeed, if the Gaussian approximation scheme is based on a first order Taylor expansion, then the mean $\mu_t$
of the ``artificial prior'' is simply the linearization point for the Taylor expansion. It is easy to check that in this case \eqref{eq:extensions-Ax}--\eqref{eq:extensions-Qx}
reduces to: $\A^{x}_t = \frac{\partial f }{\partial z_t} (u_t, \mu_t, 0)$, $\f^{x}_t = f(u_t, \mu_t, 0) - \A^{x}_t \mu_t$,
and $Q^{x}_t = \frac{\partial f }{ \partial v_t }(u_t, \mu_t, 0) \frac{ \partial f }{ \partial v_t} (u_t, \mu_t, 0)^\+$.
Hence, in this case the results will indeed be independent of the covariance matrix $\Sigma^\ls_t$
of the artificial prior in~\eqref{eq:extensions_approx_gammadef}, and choosing $\gamma_t(\ls_t)$ is equivalent
to choosing the linearization point $\mu_t$.



\subsection{MCMC and particle rejuvenation}
The \ffbs algorithm \cite{GodsillDW:2004} forms the basis for the proposed \rbps.
It has been recognized that two shortcomings of this algorithm are:
\emph{(i)} its computational complexity is of order $\Ordo(\Np\Mp\T)$, which can sometimes be prohibitively
large, and \emph{(ii)} the states simulated in the backward pass are constrained to the
support of the forward filter particles. However, in \cite{BunchG:2013} a
modification of \ffbs which addresses both of these issues is proposed.
The idea is to make use of Markov chain Monte Carlo (\mcmc) within the backward simulator to generate the backward
trajectories---the same technique can be used also with the proposed \rbps, as we discuss below.

As before, let $\bnls_{t+1:\T}$ be a partial backward trajectory. To extend this
trajectory to time $t$, instead of simulating $\nls_{1:t}$ from the backward kernel
approximation \eqref{eq:rbps_bwdkernel_pmapprox}, we draw from some \mcmc
kernel which leaves the backward kernel invariant.
Following \cite{BunchG:2013}, we use the \rbpf particles, not from time $t$, but
from time $t-1$, to define the \mcmc proposal:
\begin{align*}
  q(\nls_{1:t} \mid \bnls_{t+1:\T}, y_{1:\T})
  \eqdef \sum_{i=1}^\Np \bpw_{t-1}^i \thinspace q(\nls_t \mid \nls_{1:t-1}^i, \bnls_{t+1:\T}, y_{1:\T}) \thinspace \delta_{\nls_{1:t-1}^i}(\nls_{1:t-1})   ,
\end{align*}
where $\{ \bpw_{t-1}^i \}_{i=1}^\Np$ and $q(\cdot)$ are chosen by the user
(see \cite{BunchG:2013} for suggestions on how to select these quantities).
Similarly to \eqref{eq:rbps_bwdkernel} we then factorize the target distribution as
\begin{multline*}
  p(\nls_{1:t} \mid \nls_{t+1:\T}, y_{1:\T})
  \propto p(y_{t+1:\T}, \nls_{t+1:\T} \mid \nls_{1:t}, y_{1:t}) \\
  \times p(y_t \mid \nls_{1:t}, y_{1:t-1}) p(\nls_t \mid \nls_{1:t-1}, y_{1:t-1})
  p(\nls_{1:t-1} \mid y_{1:t-1}).
\end{multline*}
Using the \rbpf particles at time $t-1$ to approximate this distribution
we obtain the Metropolis-Hastings acceptance probability for
a proposed move $\nls_{1:t}^{(r)} \rightarrow \nls_{1:t}^*$,
\begin{align*}
  \min\left\{ 1, \frac{  \widehat p^\Np(\nls_{1:t}^* \mid \bnls_{t+1:\T}, y_{1:\T}) }{   q(\nls_{1:t}^* \mid \bnls_{t+1:\T}, y_{1:\T}) }
 \frac{   q(\nls_{1:t}^{(r)} \mid \bnls_{t+1:\T}, y_{1:\T}) }{   \widehat p^\Np(\nls_{1:t}^{(r)} \mid \bnls_{t+1:\T}, y_{1:\T}) } \right\},
\end{align*}
where
\begin{multline*}
   \frac{  \widehat p^\Np(\nls_{1:t}^* \mid \bnls_{t+1:\T}, y_{1:\T}) }{   q(\nls_{1:t}^* \mid \bnls_{t+1:\T}, y_{1:\T}) }
   = \frac{  w_{t-1}^* p(\nls_t^* \mid \nls_{1:t-1}^*, y_{1:t-1}) }{ \bpw_{t-1}^* q(\nls_t^* \mid \nls_{1:t-1}^*, \bnls_{t+1:\T}, y_{1:\T}) } \\
   \times p(y_{t+1:\T}, \bnls_{t+1:\T} \mid \nls_{1:t}^*, y_{1:t}) p(y_t \mid \nls_{1:t}^*, y_{1:t-1})
\end{multline*}
and analogously for the second factor (here, $\{\nls_{1:t-1}^*, w_{t-1}^*\}$
refers to the \rbpf particle $\{\nls_{1:t-1}^i, w_{t-1}^i\}$ such that
$\nls_{1:t}^* = ( \nls_{1:t-1}^i, \nls_{t}^*)$).

Importantly, the expression above depends on the forward \rbpf particle system only through the
proposed sample $\nls_{1:t-1}^*$. Consequently, the computational
complexity of simulating each individual backward trajectory is independent of the number
of forward filter particles $\Np$. Hence, if we run the \mcmc sampler
for $\Rmcmc$ steps at each time point we get a total computational complexity of order $\Ordo(\Rmcmc \Mp \T)$. As pointed out in \cite{BunchG:2013}, $\Rmcmc$ can typically
be chosen much smaller than $\Np$, resulting in a significant reduction
in computational complexity. Furthermore, since we simulate $\nls_t^*$ from
(the possibly continuous) proposal density $q(\cdot)$, the backward trajectories
are not constrained to the support of the forward filter particles.

A related technique is to use rejection sampling to simulate the backward
trajectories, as has been proposed by \cite{DoucGMO:2011} for the \ffbs.
However, this requires an upper bound on the backward sampling weights
\eqref{eq:rbps_weights} that holds uniformly for all backward trajectories
$\{ \bnls_{t+1:\T}^j \}_{j=1}^\Mp$. It is not obvious how to choose this
bound in the Rao-Blackwellized setting, making this technique less suitable
for the \rbps.

\subsection{Square-root implementation}\label{sec:extensions:squareroot}
As pointed out in Section~\ref{sec:bkg_rbpf}, it is in general recommended to implement the conditional Kalman filter of the \rbpf
on square-root form, to ensure symmetry and positive definiteness of the involved covariance matrices.
The same holds for the conditional backward information filter. In this section, we show how to implement the
backward recursions given by \eqref{eq:rbps_hier_statistics_pred}, \eqref{eq:rbps_hier_statistics_update} and \eqref{eq:rbps_mixed_statistics_pred}
on square-root form.

We use the technique proposed by \cite{KailathSH:2000}, which is based on a numerically robust QR-factorization,
and adapt this to the present setting.
For an arbitrary matrix $A$, we can factorize it as $A = \QQ\RR$, where $\QQ$ is orthogonal and $\RR$ is upper triangular.
Let $\Omega_t^{1/2}$ be a matrix such that $\Omega_t = \Omega^{1/2}_t \Omega^{\+/2}$, and similarly for $\widehat\Omega_t$.
Rather than computing the information matrices $\Omega_t$ and $\widehat\Omega_t$ in the backward filter, we will
propagate the square-roots $\Omega^{1/2}_t$ and  $\widehat\Omega^{1/2}_t$.

Consider first the backward measurement update \eqref{eq:rbps_hier_statistics_update}. We compute a QR-factorization
of the matrix,
\begin{align}
  \begin{pmatrix}
    \Omega^{\+/2}_{t} \\ (R^{1/2}_t)^{-1} C_t
  \end{pmatrix} = \QQ
  \begin{pmatrix}
    \RR_1 \\ 0
  \end{pmatrix}.
\end{align}
Here, $R_t^{1/2}$ can be computed by a Cholesky factorization of the measurement noise covariance matrix $R_t$.
It follows that $\RR_1^\+ \RR_1 = \Omega_t + \C_t^\+ R_t^{-1} \C_t$, which implies that
$\widehat\Omega^{1/2}_t = \RR_1^\+$.

Next, we consider the backward prediction for the hierarchical \clg model, given by \eqref{eq:rbps_hier_statistics_pred}.
We compute a QR-factorization of the following matrix:
\begin{align}
  \begin{pmatrix}
    I & 0 \\ \widehat\Omega^{\+/2}_{t+1} \F_{t+1} & \widehat\Omega^{\+/2}_{t+1} \A_{t+1}
  \end{pmatrix} = \QQ
  \begin{pmatrix}
    \RR_1 & \RR_2 \\ 0 & \RR_3
  \end{pmatrix}.
\end{align}
It follows that
\begin{align}
 \label{eq:rbps_sqroot_matrix}
  \begin{pmatrix}
    \RR_1^\+ \RR_1 &  \RR_1^\+ \RR_2 \\ 
    \RR_2^\+ \RR_1 &  \RR_2^\+ \RR_2 + \RR_3^\+ \RR_3
  \end{pmatrix} 
  =
  \begin{pmatrix}
    I + \F_{t+1}^\+ \widehat\Omega_{t+1} \F_{t+1} &  \F_{t+1}^\+ \widehat\Omega_{t+1} \A_{t+1} \\
    \A_{t+1}^\+ \widehat\Omega_{t+1} \F_{t+1} &  \A_{t+1}^\+ \widehat\Omega_{t+1} \A_{t+1} \\
  \end{pmatrix}.
\end{align}
From \eqref{eq:rbps_sqroot_matrix} and \eqref{eq:rbps_hier_statistics_pred}, we can identify
\begin{subequations}
  \begin{align}
    \RR_1^\+ \RR_1 &= M_{t+1}, \\
    \RR_2^\+ &=  \A_{t+1}^\+ \widehat\Omega_{t+1} \F_{t+1} M_{t+1}^{-1} \RR_1^{\+}, \\
    \RR_3^\+ \RR_3 &= \A_{t+1}^\+ \widehat\Omega_{t+1} \A_{t+1} - \RR_2^\+ \RR_2 = \Omega_t.
  \end{align}
\end{subequations}
Hence, $\Omega^{1/2}_{t} = \RR_3^\+$ and $\lambda_t = (\A_{t+1}^\+ - \RR_2^\+ \RR_1^{-\+} \F_{t+1}^\+) \errls_{t+1}$.

Similarly, we can address the backward prediction for the mixed linear/nonlinear model \eqref{eq:rbps_mixed_statistics_pred} by
computing the QR-factorization,
\begin{align}
  \begin{pmatrix}
    I & 0 \\
    \widehat\Omega^{\+/2}_{t+1} \F_{t} \gsmatrix_{t}  & \widehat\Omega^{\+/2}_{t+1} \bar\A_{t} \\
    0 & (\Q_t^{1/2})^{-1} \bar\B_t
  \end{pmatrix} = \QQ
  \begin{pmatrix}
    \RR_1 & \RR_2 \\ 0 & \RR_3 \\ 0 & 0
  \end{pmatrix}.
\end{align}
By similar computations as above, we get
$\Omega^{1/2}_{t} = \RR_3^\+$ and $\lambda_t = (\bar\A_{t}^\+ - \RR_2^\+ \RR_1^{-\+} \gsmatrix_{t} \F_{t}^\+) \errls_{t}$.

\section{Experimental results}
We evaluate the proposed \rbps on two numerical examples
and compare its performance to alternative smoothers.
The following methods are considered:
\begin{itemize}
\item \ffbs: A non-Rao-Blackwellized \ffbs \cite{GodsillDW:2004}.
\item \rbfs: A Rao-Blackwellized Kitagawa smoother \cite{Kitagawa:1996}.
\item \rbffjbs:  Rao-Blackwellized forward filter/joint backward simulator~\cite{FongGDW:2002}.
\item \rbffbs:  The proposed method (Algorithm~\ref{alg:rbps-2}).
\end{itemize}
For all methods, a bootstrap PF \cite{GordonSS:1993} or RBPF \cite{ChenL:2000,SchonGN:2005} is used in the forward direction.

The \rbfs consists of running an RBPF and storing the nonlinear state trajectories. Smoothed linear state estimates
are then computed by running constrained Rauch-Tung-Striebel (RTS) smoothers \cite{RauchTS:1965}, conditionally on these nonlinear trajectories.
The \rbffjbs is an adaptation of the ``joint backward simulator'' by \cite{FongGDW:2002}, which
runs an \rbpf in the forward direction, but samples $(\nls_t, \ls_t)$ jointly in the backward direction.
The method relies on having access to the linear state samples in order to compute the backward sampling probabilities.
In fact, the method given in \cite{FongGDW:2002} is only applicable to hierarchical \clg models,
but we modify it to work also for mixed linear/nonlinear \clg{s}. Furthermore,
we complement the method with constrained RTS smoothing to compute refined smoothed linear state estimates,
which makes a more fair comparison (indeed, this is a simple ``trick'' that can be used to improve the performance of
the method by~\cite{FongGDW:2002}).

\subsection{Estimation of a time-varying parameter}
We consider first a 5th order mixed linear/nonlinear system. The nonlinear part is given by the time series,
\begin{subequations}
  \label{eq:numex_mln_nls}
  \begin{align}
    \nls_{t+1} &= 0.5\nls_t + \theta_t \frac{\nls_t}{1+\nls_t^2} + 8\cos(1.2t) + 0.071v_{t}^\nls, \\
    y_t &= 0.05\nls_t^2 + e_t,
  \end{align}
\end{subequations}
for some process $\process{\theta_t}$. The case with a static $\theta_t \equiv 25$ has been studied by,
among others, 
\cite{GordonSS:1993}.
 Here, we assume instead that $\theta_t$ is
a time varying parameter with known dynamics, given by the output from a 4th order linear system,
\begin{subequations}
  \label{eq:numex_mln_ls}
  \begin{align}
    \ls_{t+1} &=
    \begin{pmatrix}
      3 & -1.691 & 0.849 & -0.3201 \\ 2 & 0 & 0 & 0 \\ 0 & 1 & 0 & 0
      \\ 0 & 0 & 0.5 & 0
    \end{pmatrix}\ls_t + 0.1v_{t}^\ls \\
    \theta_t &= 25 +
    \begin{pmatrix}
      0 & 0.04 & 0.044 & 0.008
    \end{pmatrix}\ls_t,
  \end{align}
\end{subequations}
with poles in $0.8\pm0.1i$ and $0.7\pm0.05i$.
Combined, \eqref{eq:numex_mln_nls} and \eqref{eq:numex_mln_ls} is a mixed linear/nonlinear system.
The noises are assumed to be white, Gaussian and mutually independent; $v_{t}^\nls \sim \N(0, 1)$,
$v_{t}^\ls \sim \N(0, I)$ and $e_t \sim \N(0,0.1)$.


We generate \thsnd{1} batches of data from the system, each with $\T = 100$ samples.
We run the smoothers two times, first with $\Np = 300$ and then with $\Np = 30$ particles.
The backward-simulation-based methods use $\Mp = \Np/3$ backward trajectories,
based on the recommendation to set $M \lesssim N$ \cite{LindstenS:2013}.
Table~\ref{tab:results_mixed_rmse} summarizes the results, in terms of the time averaged root-mean-squared errors (RMSE) for the nonlinear
state $\nls_t$ and for the time varying parameter $\parameter_t$ (note that $\parameter_t$ is a linear combination of the
four linear states $\ls_t$).
We emphasize that the RMSE values are computed with respect to the ``true trajectories'',
and not with respect to the optimal smoother (which is intractable). That is, even the optimal smoother would have resulted
in a non-zero RMSE, and this should be taken into account when interpreting the results reported in the table.

\begin{table}[H]
  \caption{RMSE values averaged over \thsnd{1} runs}
  \centering
  \begin{tabular}{lccccc}
    \toprule
    & \multicolumn{2}{c}{$\Np = 300$} && \multicolumn{2}{c}{$\Np = 30$} \\
    Smoother & $\nls_t$ & $\parameter_{t}$ && $\nls_t$ & $\parameter_{t}$ \\
    \midrule
    \ffbs & $0.499 $ & $0.782 $ && $1.203 $ & $1.238 $ \\ 
    \rbfs & $0.424 $ & $0.660 $ && $0.980 $ & $0.909 $ \\ 
    \rbffjbs & $0.399 $ & $0.579 $ && $0.967 $ & $0.869 $ \\ 
    \rbffbs & $0.398 $ & $0.564 $ && $0.965 $ & $0.836 $ \\ 
    \bottomrule
  \end{tabular}
  \label{tab:results_mixed_rmse}
\end{table}

The proposed \rbffbs gives the most accurate results among the considered smoothers, both for $\Np = 300$ and $\Np = 30$.
The difference between \rbffbs and \rbffjbs is quite small. However, standard statistical hypothesis tests indicate indeed a
clear statistically significant improvement for \rbffbs over \rbffjbs.
In fact, the small difference is not surprising, since these two methods are similar in many respects.
We discuss this further in Section~\ref{sec:discussion}.

\begin{figure}[ptb]
  \centering
  \includegraphics[width = 0.48\columnwidth]{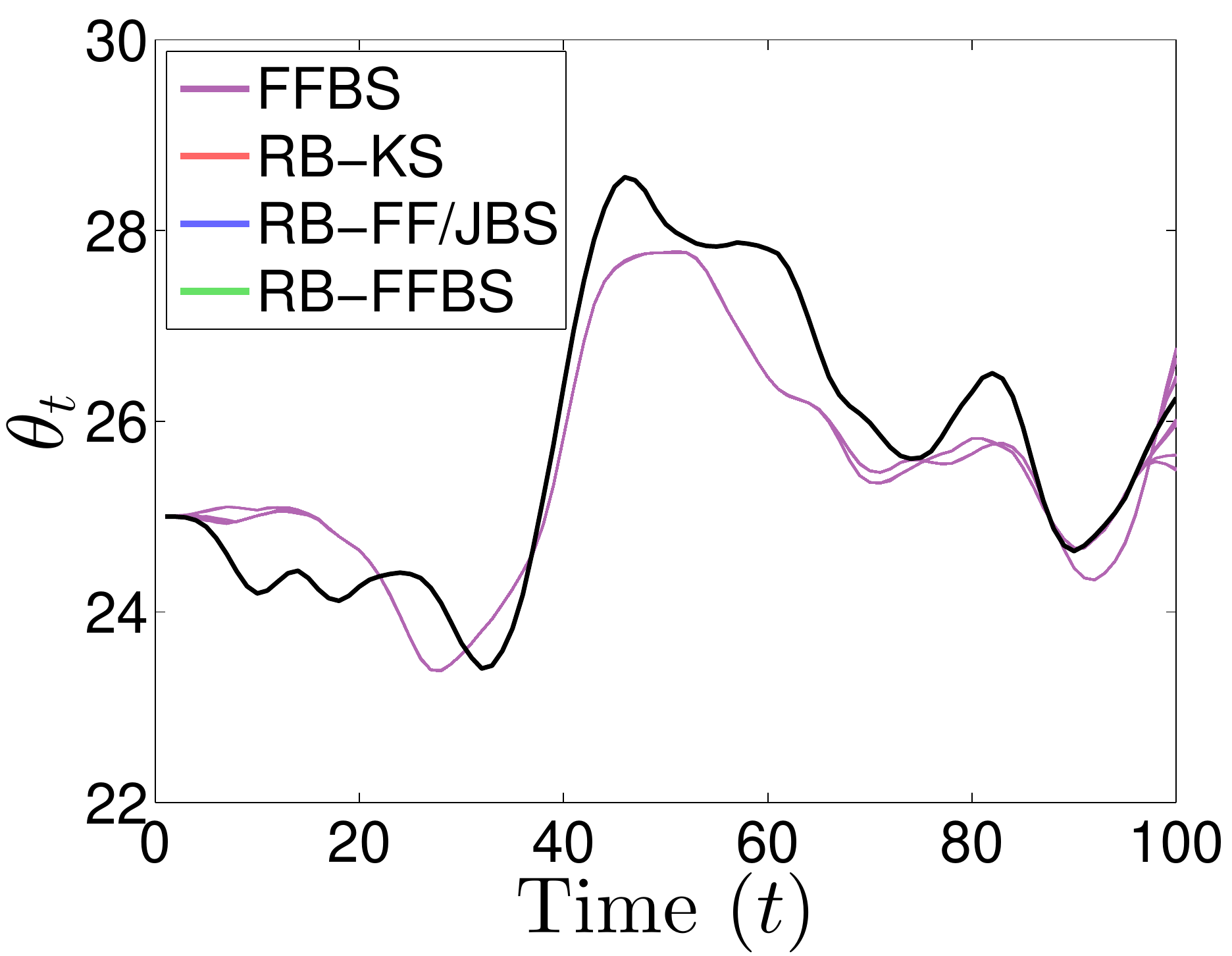}
  \includegraphics[width = 0.48\columnwidth]{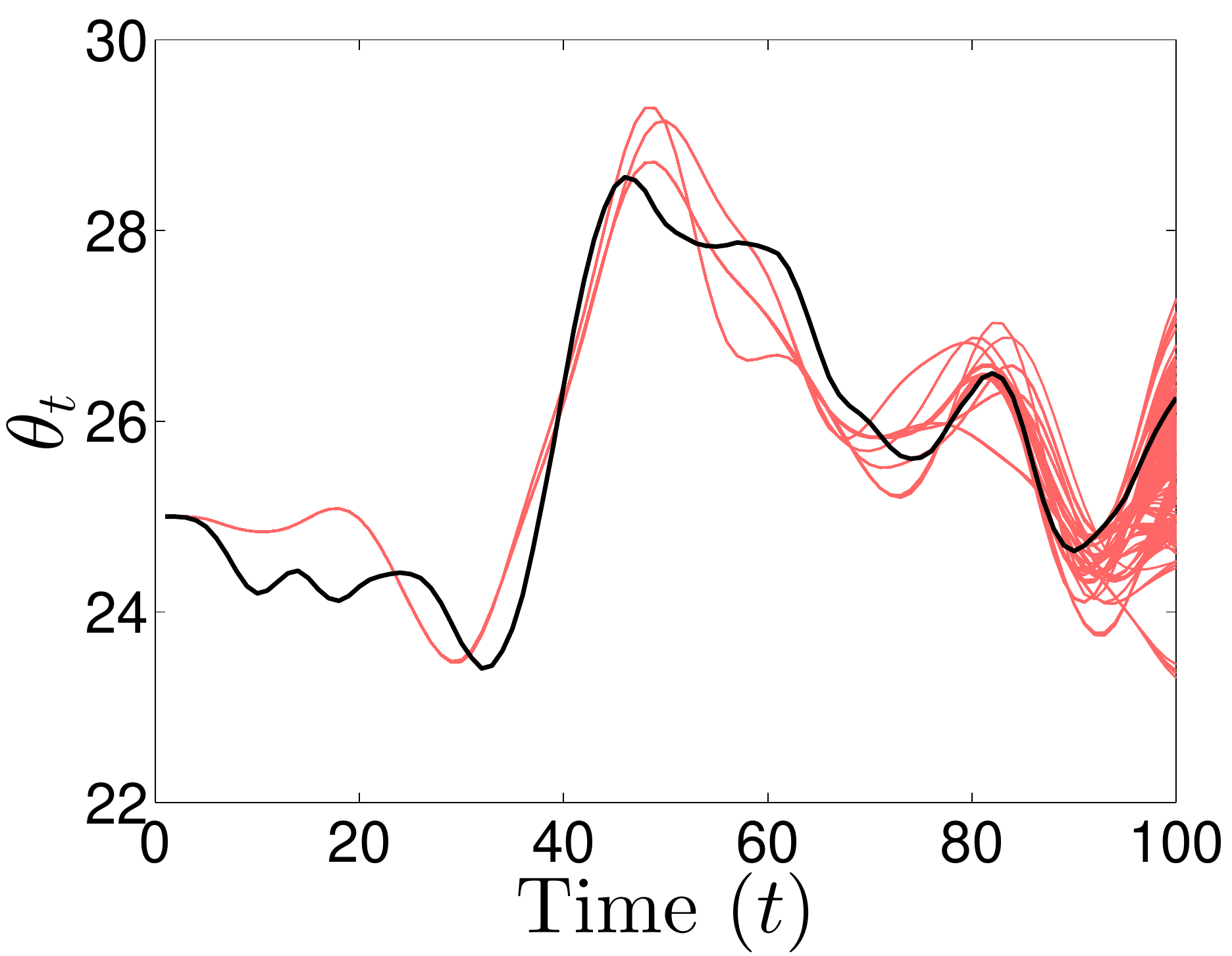}\\
  \includegraphics[width = 0.48\columnwidth]{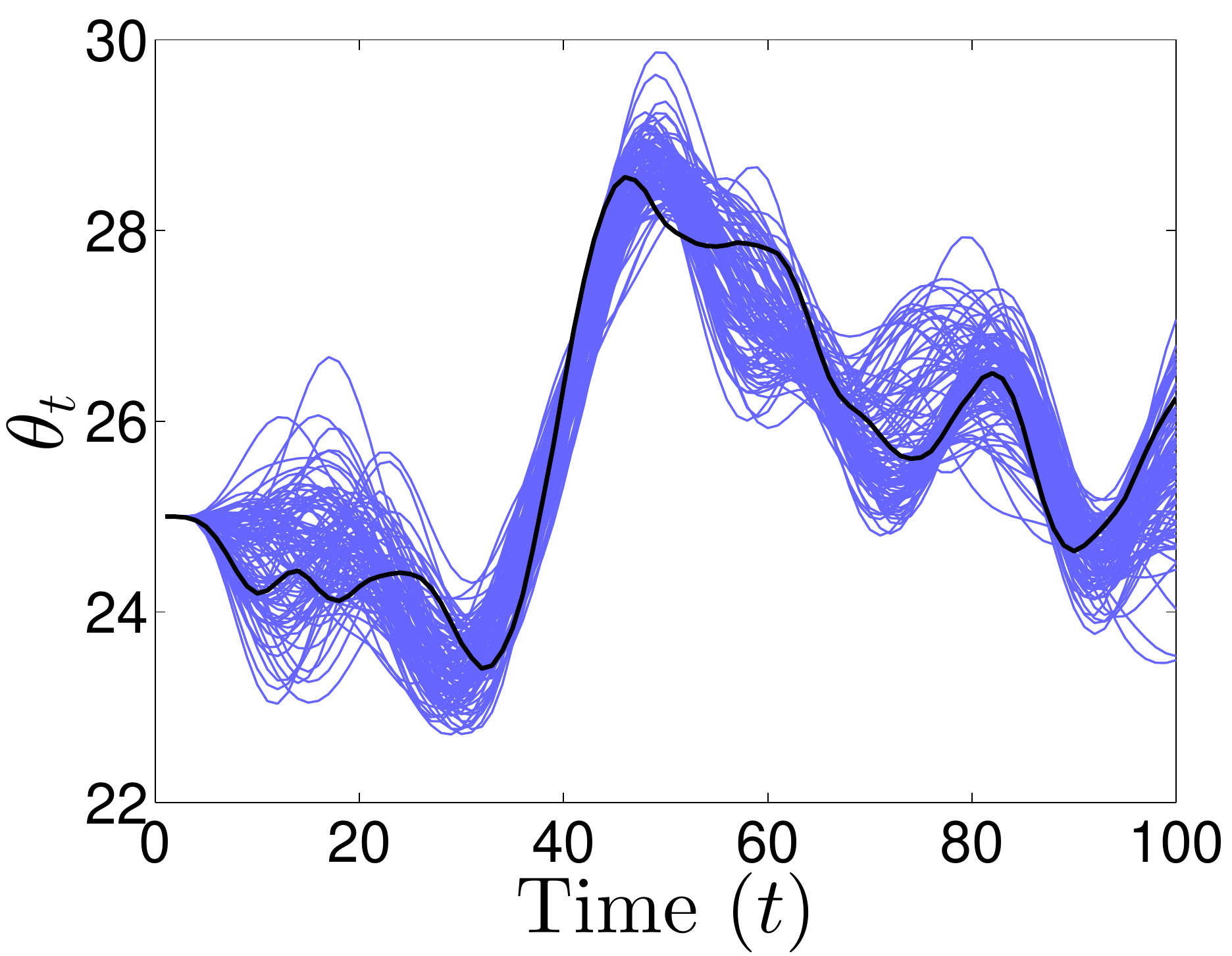}
  \includegraphics[width = 0.48\columnwidth]{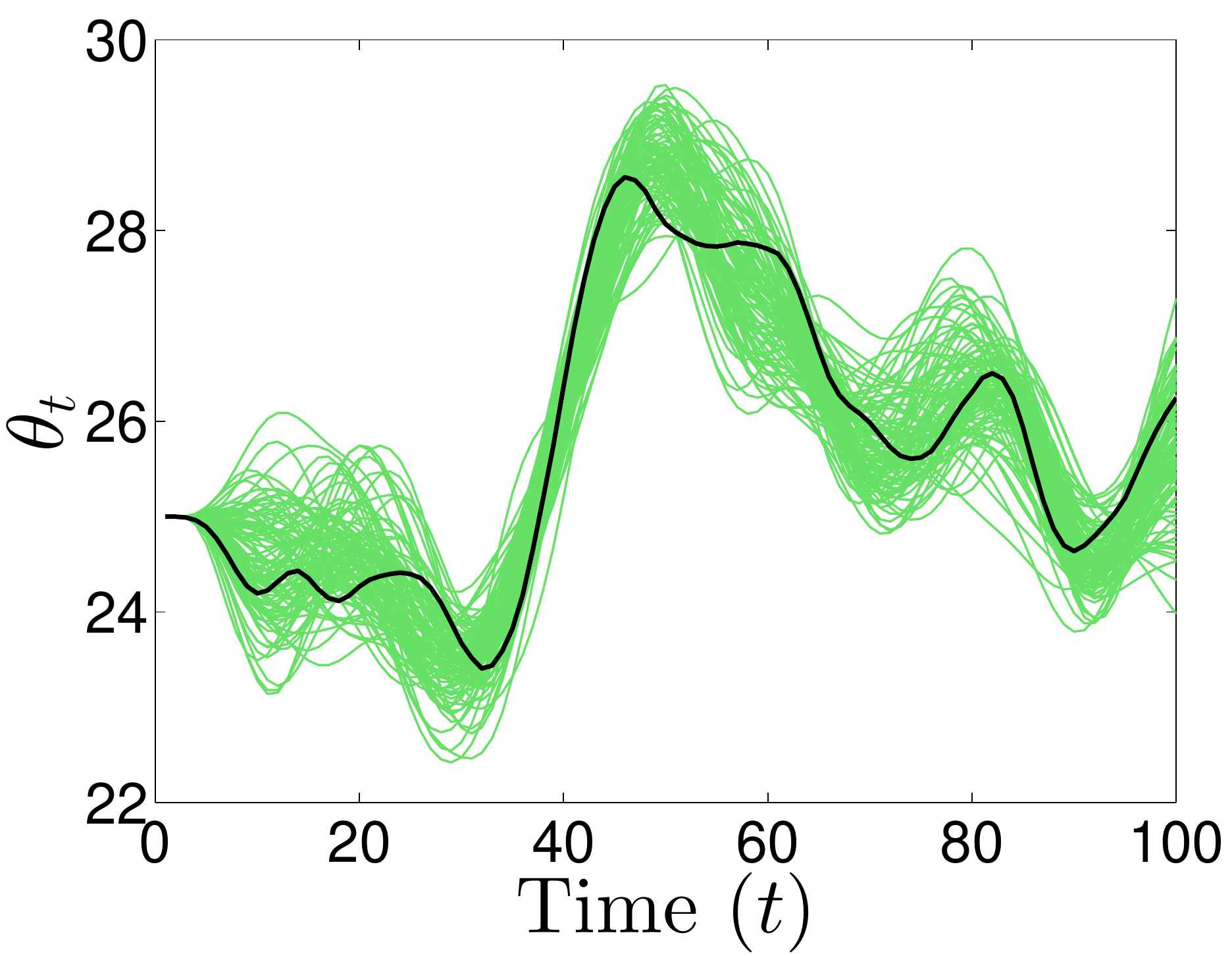}
  \caption{Estimates of $\theta_t$ for $t = \range{1}{\T}$. From top left to bottom right; \ffbs, \rbfs, \rbffjbs and \rbffbs.
    Each curve corresponds to one particle trajectory ($\widetilde \theta_{1:\T}$ for \ffbs and
    $\E[ \theta_{1:\T} \mid \widetilde \nls_{1:\T}, y_{1:\T}]$ for the Rao-Blackwellized smoothers).
    The true value is shown as a thick black line.}
  \label{fig:results_th}
\end{figure}

For further comparison, Figure~\ref{fig:results_th} shows the estimates of $\theta_t$ for one specific batch of data, using $\Np = 300$ and $\Mp = 100$.
This reveals a clear difference between the methods' abilities of accurately representing the posterior distribution of $\theta_t$.
For \ffbs and \rbfs (the top row), there is a clear degeneracy in the trajectories. For \rbfs, this is expected, as it is a direct effect of the
path degeneracy of the RBPF. For the (non-Rao-Blackwellized) \ffbs, the degeneracy is caused by the fact that $\Np = 300$ particles
is insufficient to represent the posterior in all five dimensions, resulting in that only a few particles get significantly non-zero weights.
This will cause the backward simulator to degenerate, in the sense that many backward trajectories will coincide.
The Rao-Blackwellized backward simulators (bottom row) perform much better in this respect, as there is a much larger diversity among the backward
trajectories.

\subsection{Tracking with a Constant Turn Model}

Next we consider the task of tracking a manoeuvering target from noisy observations. A two dimensional constant turn model is used (see \cite{Li2003} for details). This has a single nonlinear state which describes the instantaneous turn rate of the target, and which evolves according to a random walk,
\begin{equation}
 \nls_{t+1} = \nls_{t} + v^{\nls}_{t}     .
\end{equation}
The process noise $v^{\nls}_{t}$ is modelled as 
Cauchy distributed centered at zero. The linear state vector comprises the position and velocity of the target in Cartesian coordinates. The transitions are described by the equation,
%
\begin{equation}
 z_{t+1} = A(u_{t+1}) z_{t} + F(u_{t+1}) v^{z}_{t}     ,
\end{equation}
with $v^{z}_{t} \sim \mathcal{N}(0,\sigma_z^2)$. See \cite{Li2003} for the definitions of $A(\nls_{t+1})$ and $F(\nls_{t+1})$. Noisy, radar-style observations are made of the target range and bearing from a fixed point (the origin),
\begin{align}\label{eq:tt:obs}
 y_{t} = \begin{bmatrix} \tan^{-1}\left(\frac{z_{t,2}}{z_{t,1}}\right) & \sqrt{z_{t,1}^2+z_{t,2}^2} \end{bmatrix}^\+ + e_{t}    ,
\end{align}
%
where the
observation noises in the bearing and range measurements are white, Gaussian and mutually independent, with variances
$\sigma_b^2$ and $\sigma_r^2$, respectively.
This model cannot be Rao-Blackwellized directly, but may be treated using the approximate method of Section~\ref{sec:extensions:approx}. Specifically, we use a linearization of the observation
model \eqref{eq:tt:obs} around the filter mean. That is, we set $\mu_t = \ls_{t|t}$ in \eqref{eq:extensions_approx_gammadef}
(as pointed out in Section~\ref{sec:extensions:approx}, the resulting method is independent of
the choice of covariance matrix $\Sigma_t^\ls$ in \eqref{eq:extensions_approx_gammadef} when using a first order
Taylor expansion).

\begin{figure}
 \centering
 \includegraphics[width=0.65\columnwidth]{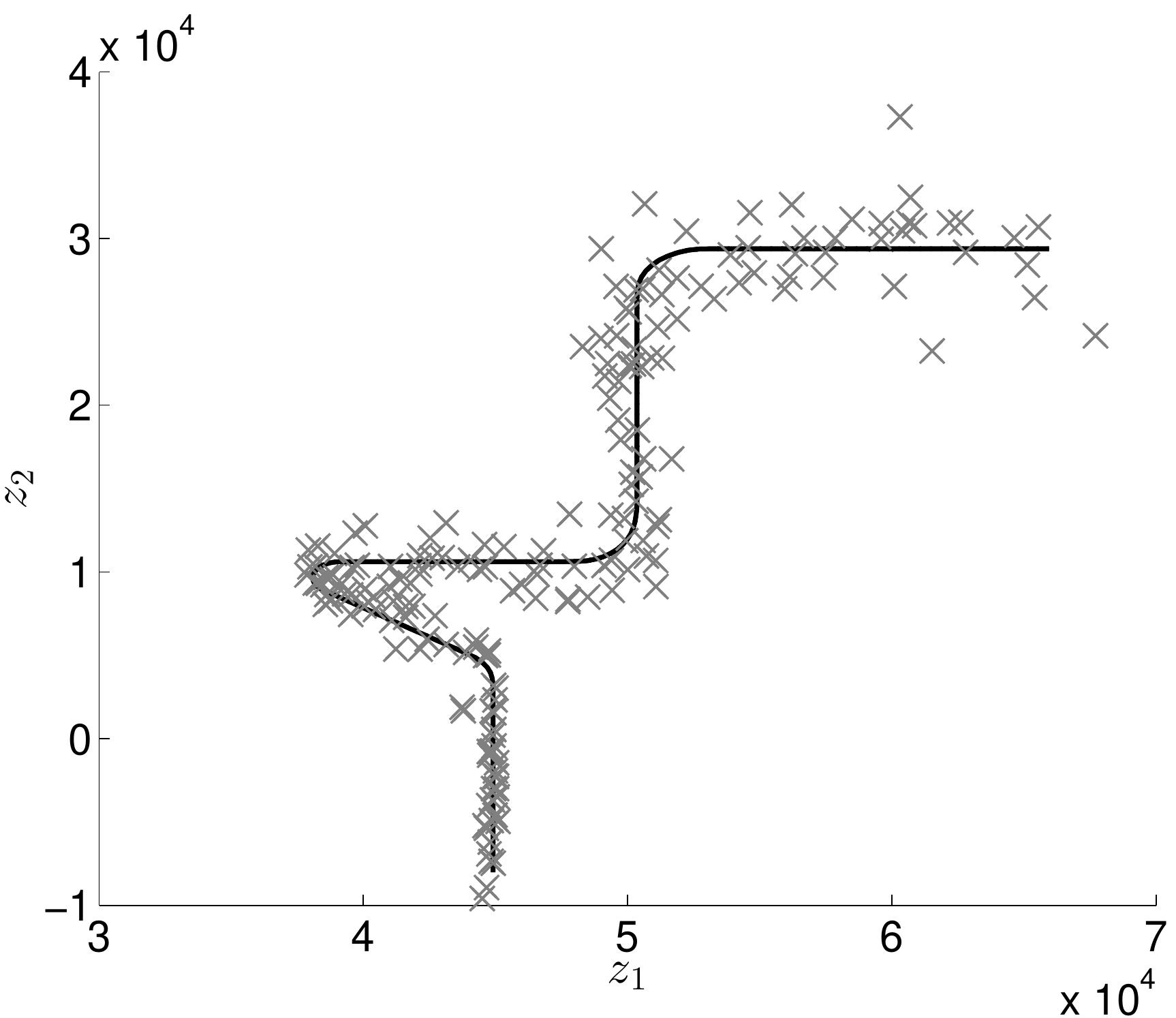}
 \caption{Benchmark fighter aeroplane trajectory from \cite{Blair1998}, with simulated observations (crosses).}
 \label{fig:trajectory}
\end{figure}

The algorithms were tested on one of the standard benchmark cases described in \cite{Blair1998}, with simulated observations made every second (see Figure~\ref{fig:trajectory}). The spread of $v^{u}_{t}$ is set to $0.03$~rad/s, the process noise standard deviation $\sigma_z = 10$~m, and the observation noise standard deviations to $\sigma_b = \frac{\pi}{90}$ and $\sigma_r = 100$~m, respectively.

We simulated 100 batches of observations. The same four algorithms were tested as for the previous model. However, the non-Rao-Blackwellized particle filter regularly failed to track the target with a reasonable number of particles, making the \ffbs impractical. It was thus excluded from the results. The approximate \rbpf used $\Np=100$ particles and the smoothers were used to sample $\Mp=100$ state sequences.

\begin{figure}
 \centering
 \includegraphics[width=0.65\columnwidth]{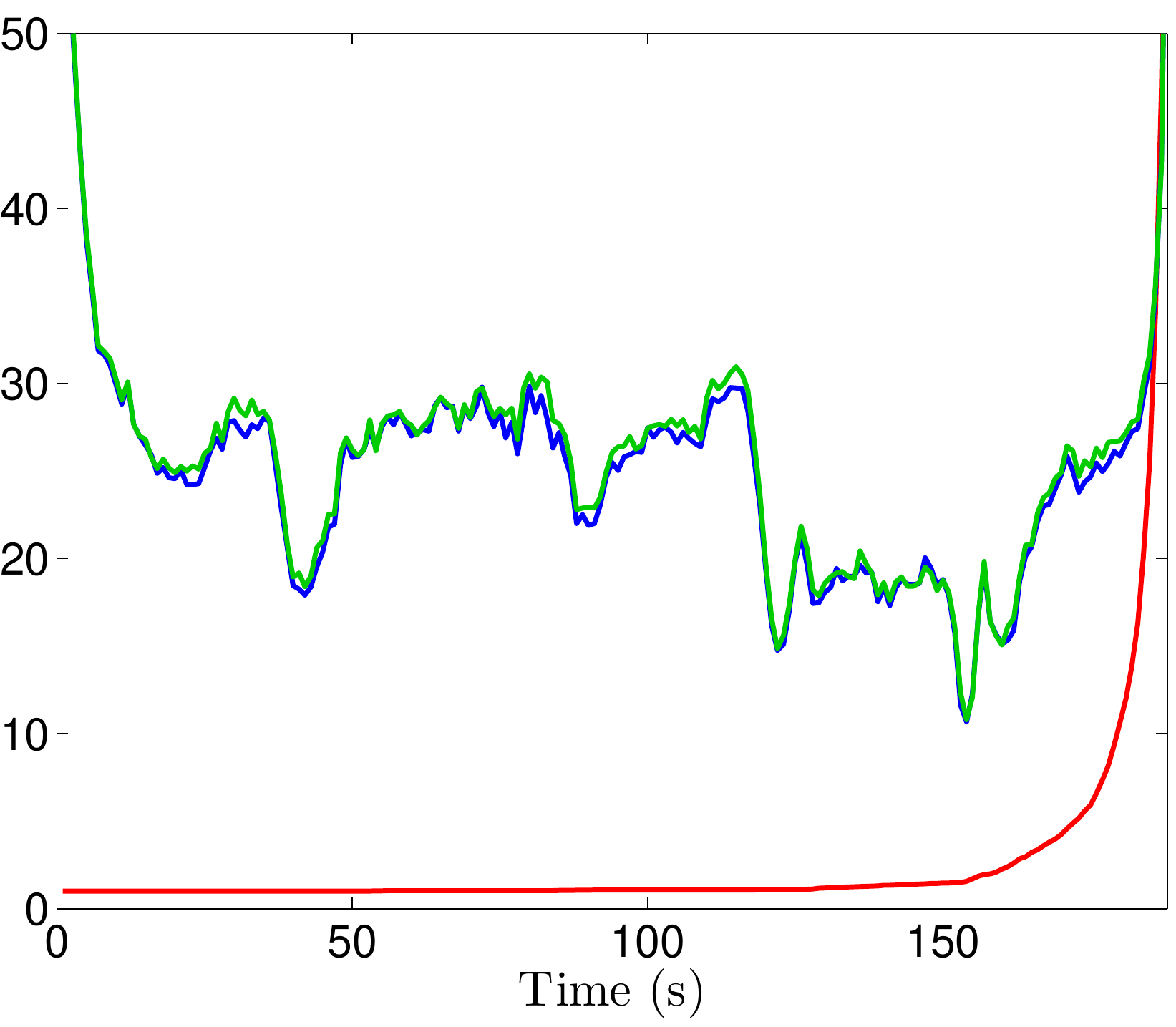}
 \caption{Average number of unique particles at each time step for \rbfs (red), \rbffjbs (blue), and \rbffbs (green).}
 \label{fig:unique_particles}
\end{figure}

\begin{figure}
 \centering
 \includegraphics[width=0.65\columnwidth]{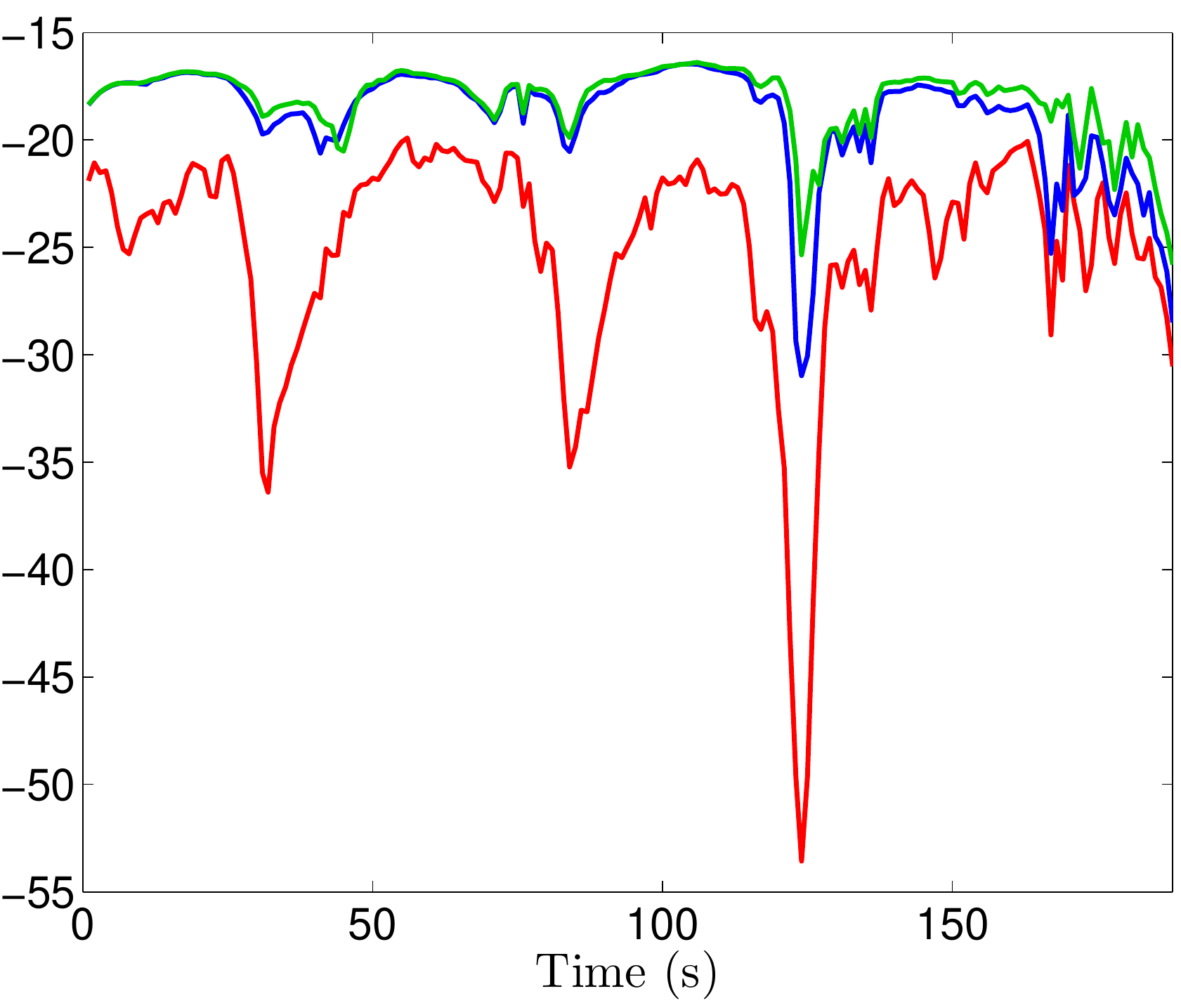}
 \caption{Average posterior log-density of the true state at each time step for \rbfs (red), \rbffjbs (blue), and \rbffbs (green).}
 \label{fig:posterior_density}
\end{figure}

RMSE values for the smoothed state estimates are shown in Table~\ref{tab:linear_rmses}.
Again, we emphasize that the RMSE values are computed \wrt to the true trajectory (Figure~\ref{fig:trajectory}),
and not \wrt to the (intractable) optimal smoother.
We see that \rbffbs gives the most accurate results.
However, the real advantage of the forward-backward smoothing algorithms is the increased number of unique particles (shown varying over time in Figure~\ref{fig:unique_particles}), which leads to a better characterisation of the posterior density. We can quantify this improvement by calculating the estimated posterior density of the true state $p(z_{t} \mid y_{1:T})$ for each approximation. This is plotted in Figure~\ref{fig:posterior_density} and clearly shows the superior performance of the forward-backward smoothing algorithms over \rbfs. Also in this respect, the \rbffbs algorithm appears to perform slightly
better than the \rbffjbs.

The experiment was repeated with different model parameters and numbers of particles. Qualitatively similar results were observed.

\begin{table}[H]
  \caption{RMSE values averaged over 100 runs}
  \centering
  \begin{tabular}{lcc}
    \toprule
    Smoother & $\nls_t$ & $\ls_t$ \\
    \midrule
    \rbfs    & 0.176 & 438 \\
    \rbffjbs & 0.152 & 382 \\
    \rbffbs  & 0.129 & 370 \\
    \bottomrule
  \end{tabular}
  \label{tab:linear_rmses}
\end{table}








\section{Discussion}\label{sec:discussion}
We have derived, within a unified framework, an \rbps
for two commonly encountered classes of conditionally linear Gaussian models; hierarchical
\clg models and mixed linear/nonlinear \clg models, respectively.
The method provides a solution to the offline (batch) state-inference problem.
Furthermore, it can be combined with standard techniques, such as particle expectation maximization 
\cite{CappeMR:2005,OlssonDCM:2008}
and particle \mcmc \cite{AndrieuDH:2010}
to address the system identification problem for these model classes
(see \cite{SvenssonLS:2014} and \cite{WhiteleyAD:2010} for these two approaches, respectively,
applied to jump Markov systems).
Compared to previously proposed \rbps, the proposed method differ on two key aspects:
\emph{(i)} it does not require any structural approximations of the model, and \emph{(ii)}
it Rao-Blackwellizes the linear state both in the forward direction and it the backward direction.

The second point is in contrast with the \rbffjbs \cite{FongGDW:2002},
 in which both the
nonlinear and the linear states are simulated in  the backward direction.
Numerically, we found that the \rbffjbs performed quite similarly to the fully Rao-Blackwellized smoother
(although, with a clear statistically significant difference in favour of the proposed method).
This is not that surprising, since, essentially, the only difference between the methods
is that for \rbffjbs the backward simulation weights are random
(they depend on the linear state samples). This gives rise to unnecessary Monte Carlo variance
which slightly deteriorates the performance of the method.
 In all other respects the two smoothers are very similar;
in particular, they make use of the same forward \rbpf to approximate the backward kernel.
In terms of computational and implementation complexity they are almost identical.
In fact, the \rbffjbs can be seen as an (unnecessary) approximation of the method proposed herein---this
approximation makes the derivation, but not the implementation or execution of the algorithm, simpler.
With this in mind we believe that the proposed \rbps indeed is the preferred method of choice of these two smoothers.
Furthermore, in our opinion, the proposed method makes use of a more intuitively correct Rao-Blackwellization,
since the marginalization is done both in the forward direction and in the backward direction.


\bibliographystyle{plain}
\bibliography{references}

\end{document}